\documentclass[subscriptcorrection,upint,varvw,barcolor=Goldenrod3,balance,hyphenate,french,pdf-a]{asmejour}

\graphicspath{{./figures/}}
\usepackage{physics}
\usepackage{newtxtext,newtxmath}
\usepackage{subcaption,float}
\usepackage[skip=0.333\baselineskip]{caption}
\usepackage{enumitem}
\usepackage{xcolor}
\usepackage[most]{tcolorbox}

\hypersetup{
	pdfauthor={John H. Lienhard},
	pdftitle={ASME Journal Paper LaTeX Template},
	pdfkeywords={ASME journal paper, LaTeX template, BibTeX style, asmejour class},
	pdfsubject = {Describes the asmejour LaTeX template},
	pdflicenseurl={https://ctan.org/pkg/asmejour},
}

\JourName{Applied Mechanics}

\begin{document}

\SetAuthorBlock{G. R. Krishna Chand Avatar\CorrespondingAuthor}{
Department of Aerospace Engineering,\\
Indian Institute of Science,\\
Bengaluru, India\\
Email: krishnaagr@iisc.ac.in
}

\SetAuthorBlock{Vivekanand Dabade \CorrespondingAuthor}{
Department of Aerospace Engineering,\\
Indian Institute of Science,\\
Bengaluru, India\\
Email: dabade@iisc.ac.in
}

\title{Kirchhoff's analogy for a planar deformable ferromagnetic rod}


\begin{abstract}
Kirchhoff's kinetic analogy relates the equilibrium solutions of an elastic rod or strip to the motion of a spinning top. In this analogy, time is replaced by the arc length parameter in the phase portrait to determine the equilibrium configurations of the rod. Predicted equilibrium solutions from the phase portrait for specific boundary value problems, as well as certain localized solutions, have been experimentally observed. In this study, we employ the kinetic analogy to investigate the equilibrium solutions of planar soft ferromagnetic rods subjected to transverse and longitudinal external magnetic fields. Our analysis reveals a subcritical pitchfork bifurcation in the phase portrait of a ferromagnetic rod subjected to transverse external magnetic field as the axial load is decreased continuously from a large compressive load. Similarly, a supercritical pitchfork bifurcation is observed in the case of longitudinal external magnetic field. We predict equilibrium configurations for a free-standing soft ferromagnetic elastic rod and the same subjected to canonical boundary conditions. Furthermore, we observe novel localized equilibrium solutions arising from homoclinic and heteroclinic orbits, which are absent in the phase portraits of purely elastic rods.
\end{abstract}

\date{Version \versionno, \today}

\maketitle

\section{Introduction}
The Kirchhoff kinetic analogy provides a rigorous mathematical framework that establishes a correspondence between three seemingly distinct physical systems: the motion of symmetric spinning tops, the dynamics of planar pendulums, and the equilibrium configurations of inextensible elastic rods \cite{vanderHeijden2000helical}. This analogy is particularly significant because it enables the transfer of well-characterized solutions and analytical techniques developed for spinning tops and pendulums to the study of elastic rods. This analogy further motivates the study of equilibrium solutions of ferromagnetic elastic rods using phase portraits, albeit through numerical integration. By mapping the governing equations of one system onto another, the analogy provides a deeper understanding of the deformation of free-standing ferromagnetic elastic rods as well as ferromagnetic elastic rods subjected to various canonical boundary conditions.

There is, however, a fundamental distinction between the dynamics of spinning tops (or planar pendulums) and the stationary solutions described by the Kirchhoff rod model. The motion of spinning tops and pendulums is governed by initial value problems, whereas the Kirchhoff rod model is formulated as a boundary value problem. For spinning tops, solutions are determined by specifying initial conditions, such as the position and velocity of the rigid body at a given time. In contrast, for elastic rods, prescribing the slope or curvature (displacement boundary conditions) and the force or moment (traction boundary conditions) at a single point does not represent a physically realistic scenario. Typically, the focus is on determining the equilibrium shape of a rod constrained at two distinct points in space.

Furthermore, in the theory of elastic rods, localized solutions constitute some of the most intriguing types of solutions. These solutions are crucial for understanding buckled instabilities in rods, where they can twist and form self-contact at discrete points. Such instabilities are observed in the helical supercoiling of DNA molecules, leading to the formation of structures known as plectonemes \cite{Clauvelin2009}. Similar self-contact localized solutions have also been reported in rods subjected to simultaneous twist and tension \cite{vanderHeijden2000helical}. To investigate localized solutions for a ferromagnetic rod, we analyze the homoclinic and heteroclinic trajectories, an example of such a solution for a symmetric rod is shown in Fig. \ref{fig:localised-longitudinal-kdbar-100}. In contrast, homoclinic solutions are not typically observed in spinning tops or pendulums, as they connect to unstable equilibrium fixed points, rendering them practically unobservable in these systems.

The motivation for this study originates from earlier work by Prof. Richard James \cite{Fosdick1981}, where the authors investigated the pure bending problem of an elastica composed of a material with a non-convex stored energy function. Their analysis revealed that, for certain terminal bending moments, the equilibrium configuration is not unique. Functional materials such as shape memory alloys typically exhibit non-convex stored energy functions. Another class of functional materials that has been extensively studied using the variational principles of micromagnetics is ferromagnetic materials \cite{james1990frustration, dabade2019micromagnetics}. This leads to a natural question: How would a ferromagnetic elastica respond to terminal loads?

In this paper, we attempt to address this question using Kirchoff's kinetic analogy. We begin by deriving the total energy of the ferromagnetic rod and demonstrate that the associated Hamiltonian, derived from the energy density (Lagrangian), is independent of the arc-length parameter. This observation enables us to invoke Kirchhoff's kinetic analogy for ferromagnetic rods. Consequently, the Hamiltonian of this system is conserved along trajectories in phase space, with each trajectory representing a different deformed configuration of the ferromagnetic rod of varying length. Our analysis reveals subcritical and supercritical pitchfork bifurcations in phase space under transverse and longitudinal external magnetic fields, respectively. Furthermore, we observe novel homoclinic and heteroclinic orbits that correspond to localized solutions of the ferromagnetic elastica. Finally, we address certain boundary value problems using Kirchhoff's kinetic analogy, which involves finding trajectories that satisfy both the boundary conditions and the length constraint.

In Section 2, we introduce the total energy functional and the Lagrangian of a ferromagnetic rod, followed by the derivation of the corresponding Hamiltonian. Section 3 is devoted to constructing the phase portrait of the Hamiltonian for both transverse and longitudinal external magnetic fields. This analysis provides the equilibrium configurations of various free-standing ferromagnetic rods, as described in Section 4. Each equilibrium configuration is determined by the selected trajectory. In Section 5, we address certain canonical boundary value problems using the phase portrait framework, under the assumption that the vertical reaction force is zero. Finally, the paper concludes with a summary and discussion presented in Section 6.

\section{Energy formulation}\label{sec:2}
In this section, we present the total energy of the ferromagnetic rod along with its energy density, represented by the Lagrangian $f$. The associated Hamiltonian is derived using the Legendre transform of $f$. We will show that the associated Hamiltonian is independent of the arc length $s$. We begin with the kinematic description of ferromagnetic rods. For a detailed similar exposition for ferromagnetic ribbons, we refer the reader to \cite{AvatarDabade2024}.

\subsection{Kinematics}
\begin{figure}[h!]
	\centering
	\includegraphics[width=0.8\linewidth]{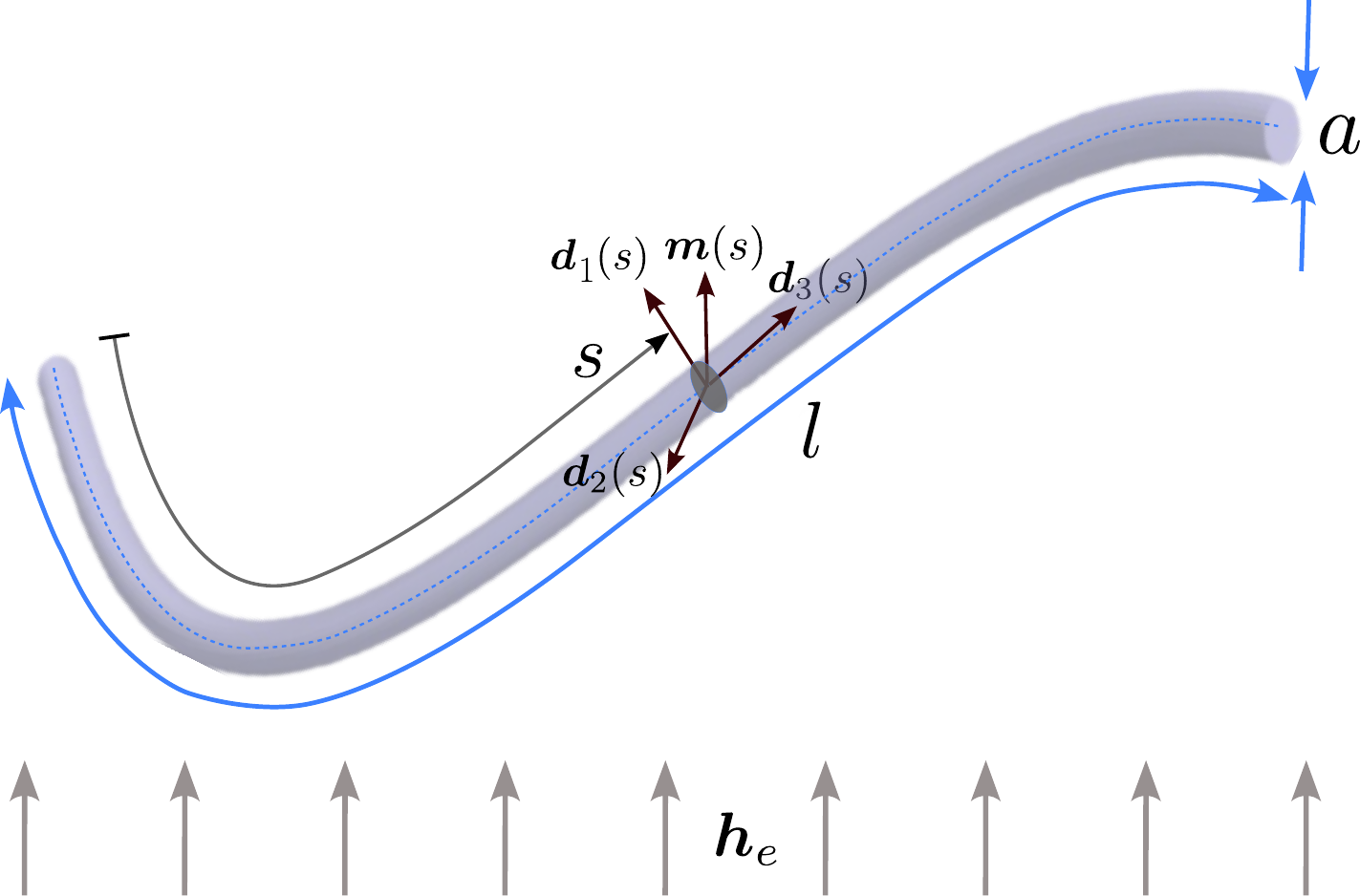}
	\caption{Geometry of the ferromagnetic rod illustrating the representation of magnetisation vector $\vb*{m}(s)$, and the material frame basis vectors $\{\vb*{d}_1(s),\vb*{d}_2(s),\vb*{d}_3(s)\}$.}
	\label{fig:rods}
\end{figure}

Rods possess two widely separated length scales—length ($l$) and diameter ($a$)—such that $a \ll l$ (see Fig. \ref{fig:rods}). In the undeformed configuration, the rod with a circular cross-section is represented in the standard Cartesian basis $(\vb*{e}_1, \vb*{e}_2, \vb*{e}_3)$ as
\begin{equation}
\vb*{X}(s, \tilde{a}_1, \tilde{a}_2) = s\vb*{e}_3 + \tilde{a}_1\vb*{e}_1 + \tilde{a}_2\vb*{e}_2,
\end{equation}
where $s$ denotes the centerline arc length coordinate, and $\tilde{a}_1$ and $\tilde{a}_2$ are the transverse coordinates.  Here, $s \in [0, l]$, $\tilde{a}_i \in \left[-\frac{a}{2}, \frac{a}{2}\right], ~i=1,2$, such that $\tilde{a}_1^2 + \tilde{a}_2^2 = \tilde{a}^2$, $\tilde{a} \in \left[0, \frac{a}{2}\right]$. The centerline representation of the deformed planar rod is given by
\begin{equation} \vb*{x}(s, \tilde{a}_1, \tilde{a}_2) = \vb*{r}(s) + \tilde{a}_1\vb*{d}_1(s) + \tilde{a}_2\vb*{d}_2(s), \end{equation}
where $\vb*{r}(s)$ denotes the position vector of a centerline material point, and $(\vb*{d}_1(s), \vb*{d}_2(s), \vb*{d}_3(s))$ is the orthonormal material frame basis. The rod is uniform and is assumed to be (i) unshearable, which implies that the cross-sections normal to the centerline in the undeformed state remain normal after deformation \cite{Bigoni2015,Audoly2010elasticity} such that $\dv{\vb*{r}(s)}{s}\cdot \vb*{d}_i(s) = 0,~i=1,2$, and (ii) inextensible, that is, $\dv{\vb*{r}(s)}{s} = \vb*{d}_3(s)$. We employ Kirchhoff's rod theory to model the deformation of the rod.

We assume the centerline to be smooth and regular, with $a < \frac{1}{\min_s\kappa(s)}$ \cite{Slastikov2011}, where $\kappa(s)$ is the curvature at $s$ \cite[Section 2.1,][]{Pressley2010book}. We also assume that the rod centerline lies in the $\vb*{e}_2-\vb*{e}_3$ plane and does not undergo any twist in the deformed configuration, implying that $\vb*{d}_1(s) = \vb*{e}_1$. The material basis vector $\vb*{d}_2(s) = \vb*{d}_3(s)\times\vb*{d}_1(s)$ \cite{Sano2022,AntmanBook}. Therefore, $\vb*{r}(s)$ is the only vector required to completely describe the kinematics of the rod.

\begin{figure}[h!]
	\centering
	\includegraphics[width=\linewidth]{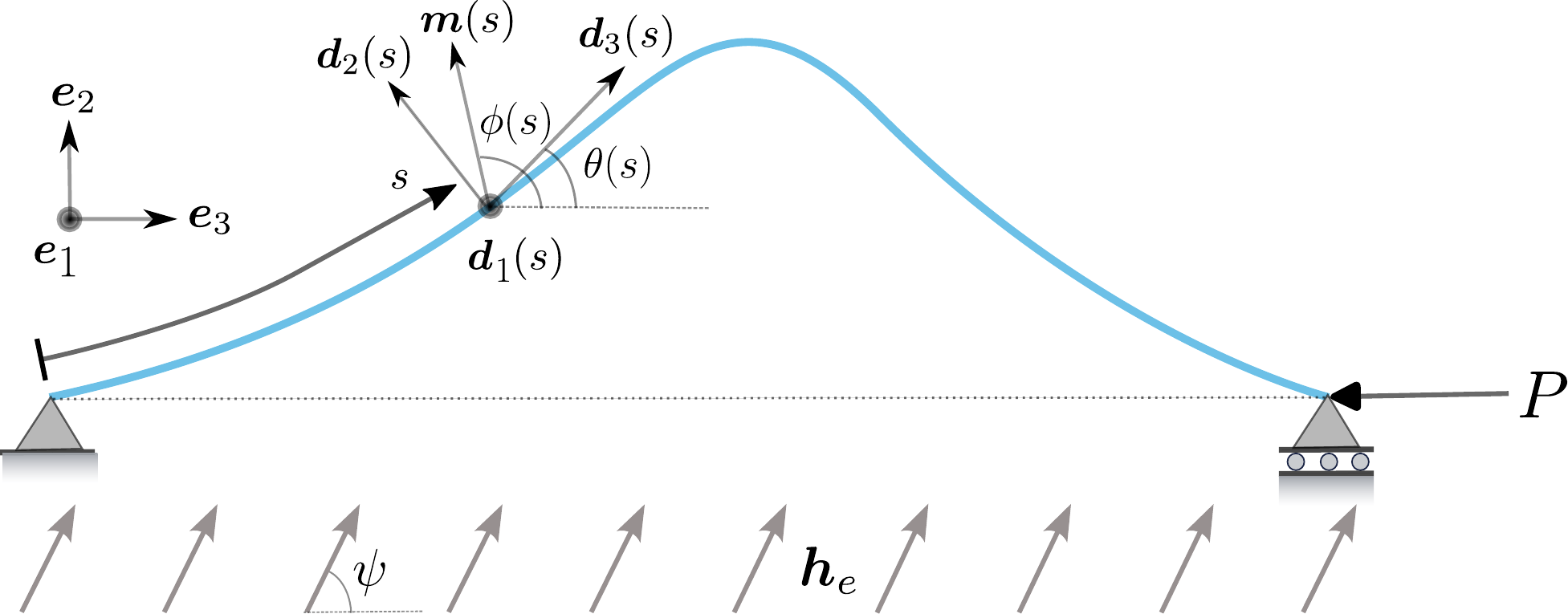}
	\caption{Schematic setup of the ferromagnetic rod.}
	\label{fig:elastica-problem-setup}
\end{figure}

The schematic of the  centerline in the deformed configuration is shown in  Fig. \ref{fig:elastica-problem-setup}. The coordinates of the centerline in $(\vb*{e}_1, \vb*{e}_2, \vb*{e}_3)$ basis are given as follows:
\begin{equation}
	x(s) = 0, ~y(s) = \int_{0}^{s} \sin\theta(\tilde s)~ d\tilde s, ~
	z(s) = \int_{0}^{s} \cos\theta(\tilde s)~ d\tilde s,     \label{eqn:arclength-parametrisation}
\end{equation}
where $\theta(s)$ denotes the angle between the tangent vector field $\vb*{d}_3(s)$ and $\vb*{e}_3$ basis vector at $s$. The tangent vector field $\vb*{d}_3(s)$ (or $\vb*{t}(s)$) is given by $\vb*{d}_3(s) = (0,\sin\theta (s),\cos\theta (s))$. Therefore, the normal vector field, as defined in \cite[Section 2.2,][]{Pressley2010book}, is $\vb*{n}(s) = \vb*{d}_2(s) =$ $\dv{\vb*{t}}{s}/\kappa(s) = (0,\cos\theta (s),-\sin\theta (s))$, where $\kappa(s)=\dv{\theta (s)}{s}$ is the bending curvature.

\subsection{Total energy}
We now construct the total free energy functional ($\mathcal{E}$) of the ferromagnetic rod, which is comprised of two components: mechanical energy ($\mathcal{E}_{\text{elastica}}$), and magnetic energy ($\mathcal{E}_{\text{magnetic}}$) such that
\begin{equation}
	\mathcal{E} = \mathcal{E}_{\text{elastica}} + \mathcal{E}_{\text{magnetic}}. \label{eqn:total-energy-functional}
\end{equation}
The mechanical energy is comprised of bending and loading device energy. The magnetic energy captures the energy associated with the magnetization of the ferromagnetic rod and its interaction with the applied magnetic field, and is motivated from the theory of micromagnetics \cite{james1990frustration,desimone2006recent}.

\subsubsection{Mechanical energy} For an inextensible rod, the primary source of elastic energy is bending energy. We denote the sum of the bending energy and the loading device energy as $\mathcal{E}_{\text{elastica}}$, which is given as follows:
\begin{align}
	\mathcal{E}_{\text{elastica}} &=  \underbrace{\frac{1}{2} \int_{0}^{l} EI \kappa^2(s) ds}_{\text{Elastic (or bending) energy}} +  \underbrace{P \left(1 - \int_{0}^{l} \cos\theta(s) ds\right),}_{\text{Loading energy}} \label{eqn:E-elastica}
\end{align}
subject to the integral constraint:
\begin{equation}
	\begin{split}
		y(l) = \int_{0}^{l} \sin\theta(s)~ ds = 0.
	\end{split}
	\label{eqn:constraints}
\end{equation}
This constraint implies that the $y$-coordinate of centerline at $s=l$ is zero. Here, $E$ is the Young's modulus, $I = \frac{\pi a^4}{64}$ is the area moment of inertia of the circular cross-section, $EI$ is the bending stiffness of the cross-section, and $P$ is the horizontal dead load (see Fig. \ref{fig:elastica-problem-setup}). We shall assume the following convention: $\bar{P}>0$ indicates compression and $\bar{P}<0$ denotes tension.

\subsubsection{Magnetic energy}
We employ the variational framework of micromagnetics to formulate the magnetic energy of a ferromagnetic rod composed of materials such as iron or nickel. The magnetisation vector, $\vb*{m}(\vb*{x})$, is the primary variable in the micromagnetics functional,and is supported on $\Omega$, where $|\vb*{m}(\vb*{x})|=1$ for $\vb*{x} \in \Omega$. Here, $\Omega$ represents the deformed planar ferromagnetic rod.

The micromagnetic energy functional \cite{dabade2019micromagnetics} comprises of the following components
\begin{multline}
	\mathcal{E}_{\text{magnetic}} = \underbrace{A\int_{\Omega} \abs{\nabla{\vb*{m}(\vb*{x})}}^{2}d\vb*{x}}_{\mathcal{E}_{\text{ex}}} + \underbrace{K_a \int_{\Omega} \Phi(\vb*{m}) d\vb*{x}}_{\mathcal{E}_{\text{anisotropy}}} + \\ \underbrace{K_d\int_{\mathbb{R}^3} \abs{\vb*{h}_m(\vb*{x})}^2~d\vb*{x}}_{\mathcal{E}_{\text{demag.}}} - \underbrace{2K_d \int_{\Omega} \vb*{h}_e \cdot \vb*{m} d\vb*{x}}_{\mathcal{E}_{\text{Zeeman}}}, \label{eqn:micromagnetics-functional}
\end{multline}
where $\mathcal{E}_{\text{ex}}$ denotes the exchange energy, $\mathcal{E}_{\text{anisotropy}}$ represents the magnetocrystalline anisotropy energy, $\mathcal{E}_{\text{demag.}}$ refers to the magnetostatic or demagnetizing (demag.) energy, and $\mathcal{E}_{\text{Zeeman}}$ is the Zeeman energy. Here, $A$, $K_{a}$ and $K_{d}$ are material parameters. The parameter $A$ is defined as the exchange constant while $K_{a}$ is known as the anisotropy constant. The magnetostatic constant $K_d$ depends on the saturation magnetization $M_s$ and is given by $K_d = \frac{M_s^2}{2\mu_0}$, where $\mu_0$ is the permeability of free space. The demagnetizing field is denoted by $\vb*{h}_m$, and the external magnetic field is represented by $\vb*{h}_e$, expressed as $\vb*{h}_e = h_e(0, \sin\psi, \cos\psi)$, where $\psi$ is the angle between $\vb*{e}_3$ and $\vb*{h}_e$ (see Fig. \ref{fig:elastica-problem-setup}).

We consider a soft ferromagnetic rod for our analysis. For soft ferromagnetic rod,  $K_{a}\rightarrow 0$, and therefore $\mathcal{E}_{\text{anisotropy}}$ is negligibly small. The magnetisation $\vb*{m}$ has no preferred direction of orientation, and aligns parallel to the sufficiently large external magnetic field $\vb*{h}_e$. Thus $\vb*{m}$ is constant. As a result, the exchange energy evaluates to zero while the magnetocrystalline energy is constant \cite{AvatarDabade2024}. Therefore, the micromagnetic energy is composed of the sum of demagnetizing and Zeeman energies. The Zeeman energy favors the alignment of $\vb*{m}$ parallel to the external magnetic field $\vb*{h}_e$, and it is expressed as follows:
\begin{equation}
	\mathcal{E}_{\text{Zeeman}} = -2K_d \frac{\pi a^2}{4} \int_{0}^{l} \vb*{h}_e \cdot \vb*{m} ds. \label{eqn:zeeman-energy-functional}
\end{equation}

For sufficiently large and fixed $\vb*{h}_e$, the magnetization vector $\vb*{m}$ remains constant, ensuring that $\vb*{h}_e\cdot\vb*{m}$ is also constant. The demag. energy is derived using a reduced magnetic energy functional as presented in \cite{Slastikov2011}, and it assumes the following form:
\begin{equation}
\mathcal{E}_{\text{demag.}} =  \frac{K_d\pi a^2}{4}\int_{0}^{l} (\vb*{m}(s)\cdot\vb*{d}_2(s))^2 ds. \label{eqn:demag-energy-functional}
\end{equation}

Note that the demagnetization energy, $\mathcal{E}_{\text{demag.}}$, depends on the deformation, as it explicitly depends on the normal vector $\vb*{d}_2(s)$ of the rod in its deformed configuration. The reader is referred to \cite{AvatarDabade2024} for the derivation of a similar expression for demag. energy of a ferromagnetic ribbon. Hence, the magnetic energy is given by
\begin{equation}
	\mathcal{E}_{\text{magnetic}} = \frac{K_d\pi a^2}{4}\int_{0}^{l} (\vb*{m}(s)\cdot\vb*{d}_2(s))^2 ds - \frac{K_d \pi a^2}{2} \underbrace{\int_{0}^{l} \vb*{h}_e \cdot \vb*{m} ds}_{\text{constant}}.
\end{equation}
Dropping the constant terms, we have the following expression for the magnetic energy:
\begin{equation}
	\mathcal{E}_{\text{magnetic}} = \frac{K_d\pi a^2}{4}\int_{0}^{l} (\vb*{m}(s)\cdot\vb*{d}_2(s))^2 ds. \label{eqn:magnetic-energy}
\end{equation}
\subsection{Total energy of ferromagnetic elastic rod}
Upon substituting the expressions of $\mathcal{E}_{\text{elastica}}$ (Eqn. (\ref{eqn:E-elastica})) and $\mathcal{E}_{\text{magnetic}}$ (Eqn. \ref{eqn:magnetic-energy}) in Eqn. \ref{eqn:total-energy-functional}, we obtain the expression of the total energy functional as follows:
\begin{multline}
	\mathcal{E}(\theta,\vb*{m}) = \frac{EI}{2} \int_{0}^{l} \bigg(\dv{\theta(s)}{s}\bigg)^2 ds + P \bigg(1 - \int_{0}^{l} \cos\theta(s) ds\bigg) \\ +  \frac{K_d\pi a^2}{4}\int_{0}^{l} (\vb*{m}(s)\cdot\vb*{d}_2(s))^2 ds.
\end{multline}
By incorporating the integral constraint (Eqn. \ref{eqn:constraints}) with the aid of a Lagrange multiplier $R$, we obtain the augmented energy functional for the ferromagnetic rod. The resulting functional is then non-dimensionalized as follows:
\begin{multline}
	\bar{\mathcal{E}}(\theta,\vb*{m}) = \frac{1}{2} \int_{0}^{1} \left(\theta'(\bar{s})\right)^2 d\bar{s} + \bar{K}_d \int_{0}^{1} (\vb*{m}(\bar{s})\cdot\vb*{n}(\bar{s}))^2 d\bar{s} \\  + \bar{P} \left(1 - \int_{0}^{1} \cos\theta(\bar{s}) d\bar{s}\right) - \bar{R} \int_{0}^{1} \sin\theta(\bar{s}) d\bar{s}. \label{eqn:elastica-energy-functional-non-dimensionalized}
\end{multline}
where $\bar{s} := \frac{s}{l}$ is the non-dimensional arc length coordinate. For notational simplicity, we define $<\cdot>(\bar{s}) := ~<\cdot>(s(\bar{s}))
$. Also, $\bar{K}_d$, $\bar{P}$ and $\bar{Q}$ are non-dimensional parameters defined as follows:
\begin{equation*}
	\bar{K}_d  = \frac{K_d \pi a^2/4}{EI/l^2} = \frac{16 K_d}{E}\left(\frac{l}{a}\right)^2, ~\bar{P} = \frac{P l^2}{EI}, ~ \bar{R} = \frac{R l^2}{EI}.
\end{equation*}
Here, $\bar{K}_d$ is an important non-dimensional number in our analysis, representing the ratio of demag. energy to elastic energy in the ferromagnetic rod. It depends on the material properties and the aspect ratio ($\frac{a}{l}$) of the ferromagnetic rod. The magnetization $\vb*{m}(s)$ makes an angle $\phi(s)$ with $\vb*{e}_3$ (see Fig. \ref{fig:elastica-problem-setup}).

In a soft ferromagnetic rod subjected to sufficiently large and fixed $\vb*{h}_e$, $\vb*{m}$ aligns along $\vb*{h}_e$ and we have $\phi(s) = \psi$. Therefore, the non-dimensionalised total energy functional for a soft ferromagnetic rod is expressed as follows:
\begin{equation}
	\begin{split}
		\mathcal{E}(\theta,\phi) &= \frac{1}{2}\int_{0}^{1} (\theta'(\bar{s}))^2 d\bar{s} + \bar{K}_d \int_{0}^{1} \sin^2(\phi(\bar{s})-\theta(\bar{s}))d\bar{s} \\
		&\quad - \bar{P}\bigg(1-\int_{0}^{1}\cos\theta(\bar{s}) d\bar{s}\bigg) - \bar{R} \int_{0}^{1} \sin\theta(\bar{s})d\bar{s} \\
	\implies \mathcal{E}(\theta,\phi) &= \int_{0}^{1} \bigg[\frac{1}{2}(\theta'(\bar{s}))^2 + \bar{K}_d \sin^2(\phi(\bar{s})-\theta(\bar{s})) + \bar{P}\cos\theta(\bar{s}) \\ &\quad\qquad - \bar{R} \sin\theta(\bar{s}) \bigg]d\bar{s} - \underbrace{\bar{P}}_{\text{constant}}.
	\end{split} \label{eqn:energy-functional-soft}
\end{equation}

\begin{figure}[h!]
	\centering
	\includegraphics[width=0.8\linewidth]{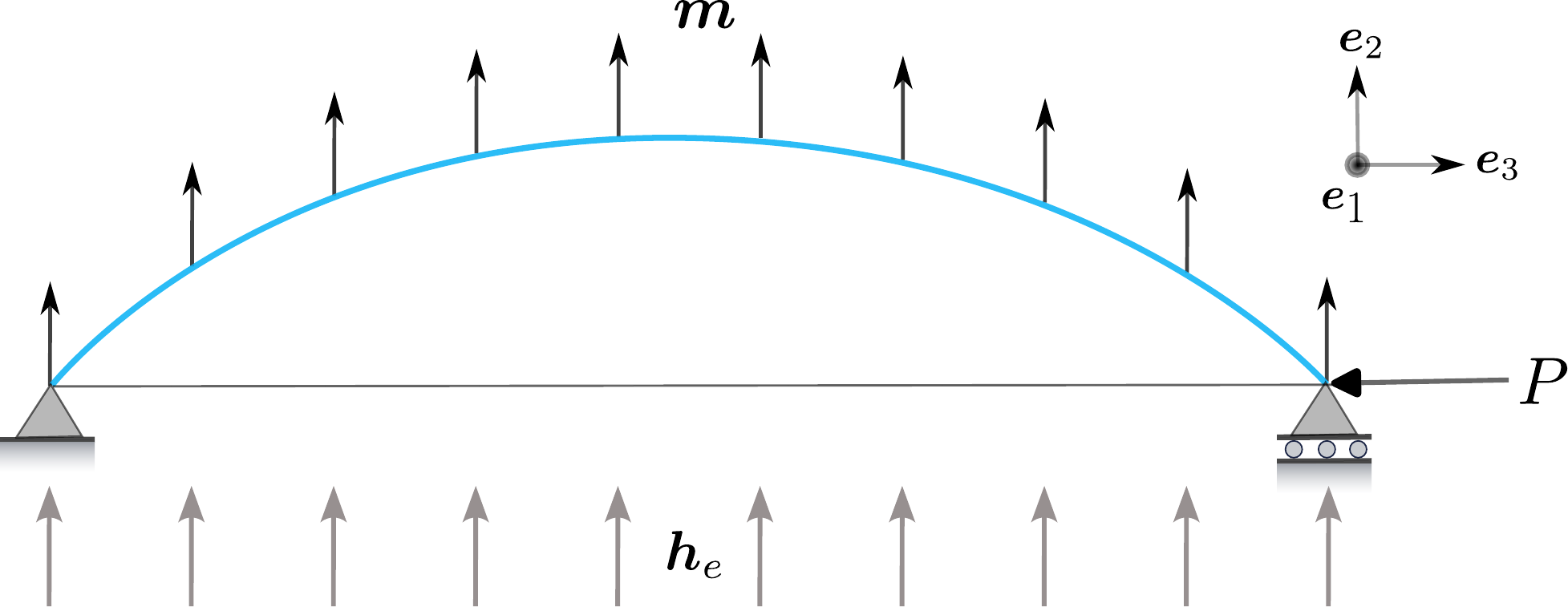}
	\caption{Soft ferromagnetic rod in the presence of a transverse $\vb*{h}_e$.}
	\label{fig:elastica-soft-magnetic-vertical-field}
\end{figure}

\begin{figure}[h!]
	\centering
	\includegraphics[width=0.8\linewidth]{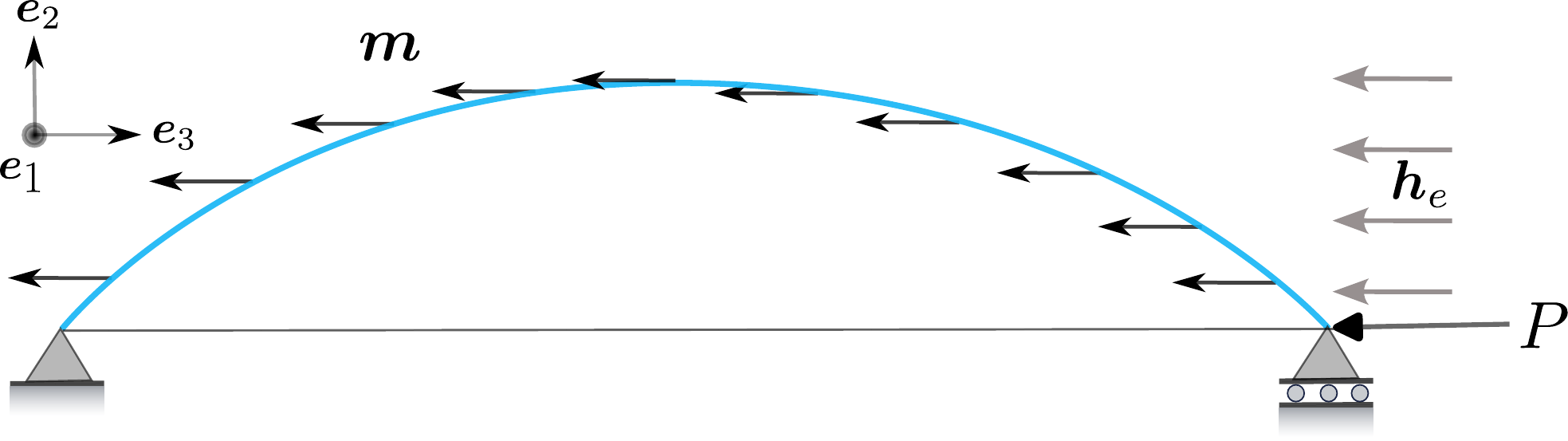}
	\caption{Soft ferromagnetic rod under the influence of a longitudinal $\vb*{h}_e$.}
	\label{fig:elastica-soft-magnetic-horizontal-field}
\end{figure}

\subsection{Hamiltonian formulation}
For a given energy density or Lagrangian $f = f(\bar{s},\theta(\bar{s}),\theta'(\bar{s}))$, the associated total energy functional is expressed as:
\begin{equation}
	I(\theta) = \int_{0}^{1} f(\bar{s},\theta(\bar{s}),\theta'(\bar{s}))d\bar{s}.
\end{equation}
We obtain the extrema of the energy functional by equating its first variational derivative to zero \cite{dacorogna2014introduction}. The extrema are the solutions to the resulting Euler-Lagrange (E-L) equation, also called the equilibrium equation, defined as:
\begin{equation}
	\dv{}{\bar{s}}\left(\pdv{f(\bar{s},\theta(\bar{s}),\theta'(\bar{s}))}{\theta'}\right) - \pdv{f(\bar{s},\theta(\bar{s}),\theta'(\bar{s}))}{\theta} = 0. \label{eqn:euler-lagrange-eqn}
\end{equation}
The Hamiltonian function ($\mathcal{H}$) is defined as the Legendre transformation of the Lagrangian ($f$), and is expressed as
\begin{equation}
	\mathcal{H}(\bar{s},\theta,v) = v\theta' - f(\bar{s},\theta,\theta').
\end{equation}
where the Hamiltonian variable  $v = \pdv{f}{\theta'}$.  For the case when $f$ is explicitly independent of $\bar{s}$, that is, $\pdv{f}{\bar{s}} = 0$, and $f = f(\theta(\bar{s}),\theta'(\bar{s}))$,
\begin{equation}
	\begin{split}
		\dv{\mathcal{H}}{\bar{s}} &= \dv{(v\theta' - f)}{\bar{s}} = \dv{}{\bar{s}}\left(\pdv{f}{\theta'}\right)\theta' + \pdv{f}{\theta'}\theta'' - \pdv{f}{\bar{s}} - \pdv{f}{\theta}\theta' - \pdv{f}{\theta'}\theta'' \\
		&= \left[\dv{}{\bar{s}}\left(\pdv{f}{\theta'}\right) - \pdv{f}{\theta}\right]\theta'.
	\end{split}
\end{equation}
Using Eqn. \ref{eqn:euler-lagrange-eqn} in the above equation gives us
\begin{equation}
	\dv{\mathcal{H}(\theta(\bar{s}),v(\bar{s}))}{\bar{s}} = 0.
\end{equation}
 Therefore, in the phase portrait of the Hamiltonian with $\theta$ and $v$ as the variables and $\bar{s}$ as the parameter, $\mathcal{H}$ remains constant as $\bar{s}$ varies along a specific trajectory. This constancy implies that:
\begin{equation}
	\mathcal{H}\left(\theta,v\right) = v\theta' - f(\theta,\theta') = \text{constant}.
\end{equation}
We specify the value of this constant, which determines the equilibrium shape of the free-standing ferromagnetic elastica. For boundary value problems, this constant is determined based on the prescribed boundary conditions.

We now proceed to derive the Hamiltonian for the soft ferromagnetic rod under two distinct cases: (i) when subjected to transverse external magnetic fields and (ii) when subjected to longitudinal external magnetic fields. The corresponding schematics for these two scenarios are illustrated in Figs. \ref{fig:elastica-soft-magnetic-vertical-field} and \ref{fig:elastica-soft-magnetic-horizontal-field}, respectively.
\subsubsection{Transverse magnetic field ($\psi = \frac{\pi}{2}$)}
Discarding the constant term, the total energy functional (Eqn. \ref{eqn:energy-functional-soft}) reduces to
\begin{multline}
	\mathcal{E}(\theta) = \int_{0}^{1} \bigg[\frac{1}{2}(\theta'(\bar{s}))^2 + \bar{K}_d \cos^2\theta(\bar{s}) + \bar{P}\cos\theta(\bar{s}) \\ - \bar{R} \sin\theta(\bar{s}) \bigg]d\bar{s}.
\end{multline}
We note that $\mathcal{E}(\theta)$ remains the same for $\psi = \frac{\pi}{2}$ and $\psi = -\frac{\pi}{2}$, this is not surprising, as all terms in the micromagnetic energy functional are even functions of the magnetization $\vb*{m}(\vb*{x})$.
The Lagrangian of the total energy functional is
\begin{multline}
	f(\theta,\theta') = \frac{1}{2}(\theta'(\bar{s}))^2 + \bar{K}_d \cos^2\theta(\bar{s}) + \bar{P}\cos\theta(\bar{s}) \\ - \bar{R} \sin\theta(\bar{s})
\end{multline}
We note that the Lagrangian is explicitly independent of the independent variable $\bar{s}$, that is, $f(\bar{s},\theta,\theta') = f(\theta,\theta')$.
The E-L equation is
\begin{equation}
	\dv{}{\bar{s}}\left(\pdv{f(\theta,\theta')}{\theta'}\right) - \pdv{f(\theta,\theta')}{\theta} = 0
\end{equation}
which implies that
\begin{equation}
	\theta''(\bar{s}) + \bar{P}\sin\theta(\bar{s}) + \bar{K}_d \sin 2\theta(\bar{s}) + \bar{R}\cos\theta(\bar{s}) = 0.
\end{equation}
In order to construct the Hamiltonian, we define the following variable $v$,
\begin{equation}
	v = \pdv{f}{\theta'} = \theta'.
\end{equation}
The Hamiltonian is obtained as
\begin{multline}
	\mathcal{H}(\theta,v) = \mathcal{H}(\theta,\theta') = \theta'v - f(\theta,\theta') = \theta'\theta' - f(\theta,\theta')  \\
	= \frac{1}{2}(\theta'(\bar{s}))^2 - \bar{K}_d \cos^2\theta(\bar{s}) - \bar{P}\cos\theta(\bar{s}) + \bar{R} \sin\theta(\bar{s})
\end{multline}
Since $\mathcal{H}(\theta, \theta')$ is constant along a trajectory, we obtain a family of curves such that
\begin{multline}
	\mathcal{H}(\theta,\theta') = \frac{1}{2}(\theta'(\bar{s}))^2 - \bar{K}_d \cos^2\theta(\bar{s}) \\ - \bar{P}\cos\theta(\bar{s}) + \bar{R} \sin\theta(\bar{s}) = C.
\end{multline}
For a free-standing elastica, $C$ can take any arbitrary value. Fig. \ref{fig:free-standing-transverse-kdbar-100} depicts the constant-Hamiltonian trajectories in the phase portrait for the transverse $\vb*{h}_e$ case, with parameters set to $\bar{P}=\frac{\pi^2}{3}$, $\bar{R}=0$ and $\bar{K}_d = 100$. Also, the figure presents the free-standing deformed shapes of a ferromagnetic rod for various specified values of $C$. In the context of boundary value problems, the constant $C$ is determined based on the prescribed boundary conditions of the planar ferromagnetic rod.

\subsubsection{Longitudinal magnetic field ($\psi = \pi$)}
The total energy functional for soft ferromagnetic rod exposed to longitudinal magnetic field is
\begin{multline}
	\mathcal{E}(\theta) = \int_{0}^{1} \bigg[\frac{1}{2}(\theta'(\bar{s}))^2 + \bar{K}_d \sin^2\theta(\bar{s}) + \bar{P}\cos\theta(\bar{s}) \\ - \bar{R} \sin\theta(\bar{s}) \bigg]d\bar{s}.
\end{multline}
For the same reason mentioned in the previous section, we observe that $\mathcal{E}(\theta)$ is identical for $\psi = 0$ and $\psi = \pi$.

The Lagrangian is  given by
\begin{multline}
	f(\theta,\theta') = \frac{1}{2}(\theta'(\bar{s}))^2 + \bar{K}_d \sin^2\theta(\bar{s}) + \bar{P}\cos\theta(\bar{s}) - \bar{R} \sin\theta(\bar{s}),
\end{multline}
and the corresponding E-L equation is as follows:
\begin{equation}
	\theta''(\bar{s}) + \bar{P}\sin\theta(\bar{s}) - \bar{K}_d \sin 2\theta(\bar{s}) + \bar{R}\cos\theta(\bar{s}) = 0.
\end{equation}
We obtain the following Hamiltonian of the problem
\begin{multline}
	\mathcal{H}(\theta,\theta') = \frac{1}{2}(\theta'(\bar{s}))^2 - \bar{K}_d \sin^2\theta(\bar{s}) \\ - \bar{P}\cos\theta(\bar{s}) + \bar{R} \sin\theta(\bar{s}) = C.
\end{multline}
Similar to the case of transverse $\vb*{h}_e$, Fig. \ref{fig:free-standing-longitudinal-kdbar-100} represents the phase portrait containing constant-Hamiltonian curves corresponding to longitudinal $\vb*{h}_e$ for the parameters set to: $\bar{P}=\frac{\pi^2}{3}$, $\bar{R}=0$ and $\bar{K}_d = 100$. The figure also presents the free-standing configurations of the ferromagnetic rods.

 The evaluation of $\bar{R}$ involves solving Eqn. \ref{eqn:constraints} along with the Hamitonian for a prescribed $C$. However, computing non-zero $\bar{R}$ is quite difficult \cite{Antman1978} and it would not be amenable to phase plane analysis. Therefore, for both transverse and longitudinal $\vb*{h}_e$ scenarios as well as purely elastic case, we study the configurations in which $\bar{R} = 0$, as explained in Section \ref{sec:bvp}.  With $\bar{R}$ unspecified, the analysis becomes more challenging. For the treatment of modes with $\bar{R} \neq 0$, we refer the reader to \cite{AvatarDabade2024}.

Therefore, we enumerate the Hamiltonian for the ferromagnetic elastic rod subjected to transverse and longitudinal $\vb*{h}_e$ as follows:
\begin{itemize}
    \item Transverse $\vb*{h}_e$:
        \begin{equation}
            \mathcal{H}(\theta,\theta') = \frac{1}{2}(\theta'(\bar{s}))^2 - \bar{K}_d \cos^2\theta(\bar{s}) \\ - \bar{P}\cos\theta(\bar{s}) = C. \label{eqn:transverse-hamiltonian}
        \end{equation}

    \item Longitudinal $\vb*{h}_e$:
        \begin{equation}
            \mathcal{H}(\theta,\theta') = \frac{1}{2}(\theta'(\bar{s}))^2 - \bar{K}_d \sin^2\theta(\bar{s}) \\ - \bar{P}\cos\theta(\bar{s}) = C. \label{eqn:longitudinal-hamiltonian}
        \end{equation}
\end{itemize}

\section{Phase portraits}
In dynamical systems theory, a phase portrait is a valuable tool for visualizing the trajectories of a dynamical system within the phase space. Phase portraits are also used to analyze critical points and their characteristics. Drawing from Kirchhoff's kinetic analogy \cite{vanderHeijden2000helical}, the phase portrait is a graphical representation of a family of constant Hamiltonian trajectories ($\mathcal{H}(\theta, \theta') = C$) in the phase plane, or the $(\theta, \theta')$-coordinate space. Here, ($\theta(\bar{s}),\theta'(\bar{s})$) denotes the coordinates of a trajectory parametrized by $\bar{s}$. The portrait is symmetric about both the $\theta'$ and $\theta$ axes since the Hamiltonian $\mathcal{H}(\theta,\theta')$  is invariant under the transformations, $\theta \rightarrow -\theta$ and $\theta' \rightarrow -\theta'$. Also, it exhibits periodicity along the $\theta$ axis with a period of $2\pi$. Due to this periodicity, we limit $\theta$ within the range $-\pi \leq \theta \leq \pi$.

We determine the critical points of the Hamiltonian and their nature. The critical points are obtained by equating the gradient of $\mathcal{H}$ to zero, that is,
\begin{equation}
	\grad \mathcal{H}  = \begin{pmatrix}
		\pdv{\mathcal{H}}{\theta} \\
		\pdv{\mathcal{H}}{\theta'}
	\end{pmatrix} = \vb*{0}.
\end{equation}
For a soft ferromagnetic rod subjected to transverse $\vb*{h}_e$, we have, using Eqn. \ref{eqn:transverse-hamiltonian},
\begin{equation}
	\begin{pmatrix}
		\pdv{\mathcal{H}}{\theta} \\
		\pdv{\mathcal{H}}{\theta'}
	\end{pmatrix} = \vb*{0} \implies \begin{cases}
		2\bar{K}_d\cos\theta(\bar{s})\sin\theta(\bar{s}) + \bar{P}\sin\theta(\bar{s}) &= 0, \\  \theta'  &= 0.
	\end{cases} \label{eqn:fixed-point-equations}
\end{equation}
For purely elastic case, we have $\bar{K}_d = 0$. Putting $\bar{K}_d = 0$ in Eqn. \ref{eqn:fixed-point-equations}, we obtain the critical points as follows:
\begin{equation}
     p_1 = (0,0), p_2 = (-\pi, 0), p_3 = (\pi,0).
\end{equation}
The critical points are indicated in the Hamiltonian phase portrait of purely elastic case in Fig. \ref{fig:section-view-elastic}. The phase portrait is generated using numerical integration and is consistent with previously reported results in the literature. This agreement enhances our confidence to proceed with the analysis of both transverse and longitudinal $\vb*{h}_e$ cases.

Additional critical points are obtained for a ferromagnetic rod exposed to magnetic fields. In the transverse $\vb*{h}_e$ case, they are obtained by solving Eqn. \ref{eqn:fixed-point-equations}:
\begin{equation}
    p_4 = \left(-\acos(-\frac{\bar{P}}{2\bar{K}_d}), 0\right),~p_5 = \left(\acos(-\frac{\bar{P}}{2\bar{K}_d}), 0\right).
\end{equation}
We indicate the critical points for transverse $\vb*{h}_e$ case in Fig. \ref{fig:section-view-transverse-kdbar-100}.

Similarly, the expressions for $p_4$ and $p_5$ for the longitudinal $\vb*{h}_e$ case (Eqn. \ref{eqn:longitudinal-hamiltonian}) are given as
\begin{equation}
	p_4 = \left(-\acos(\frac{\bar{P}}{2\bar{K}_d}), 0\right),~p_5 = \left(\acos(\frac{\bar{P}}{2\bar{K}_d}), 0\right).
\end{equation}
Fig. \ref{fig:section-view-longitudinal-kdbar-100} shows the critical points for a ferromagnetic rod subjected to longitudinal $\vb*{h}_e$.

We determine the nature of critical points using Hessian of the Hamiltonian:
\begin{equation}
	\grad^2\mathcal{H} = \begin{bmatrix}
		\frac{\partial^2\mathcal{H}}{\partial\theta^2}  & \pdv[2]{\mathcal{H}}{\theta}{\theta'} \\
		\pdv[2]{\mathcal{H}}{\theta}{\theta'} & \frac{\partial^2\mathcal{H}}{\partial\theta'^2}
	\end{bmatrix} = \begin{bmatrix}
	2\bar{K}_d \cos2\theta + \bar{P}\cos\theta & 0 \\
	0 & 1
\end{bmatrix}.
\end{equation}
The critical point is a center if $\det(\grad^2\mathcal{H}) = \frac{\partial^2\mathcal{H}}{\partial\theta^2} > 0$, and a saddle point when $\det(\grad^2\mathcal{H}) < 0$. The trajectories encircling the center are characterised by closed orbits, whereas at the saddle point, trajectories either emanate from or converge to this point. The separatrix is a non-periodic trajectory that separates the region containing closed-loop trajectories within it and the open-loop trajectories outside. It connects the unstable manifold of one saddle point to the stable manifold of another saddle point. Closed-loop trajectories are periodic in nature, and the corresponding deformed shapes exhibit inflectional points. In contrast, open-loop trajectories are non-periodic and their configurations do not possess any points of inflection.

\subsection{Purely elastic case}
\begin{figure}[h!]
	\centering
	\includegraphics[width=0.9\linewidth]{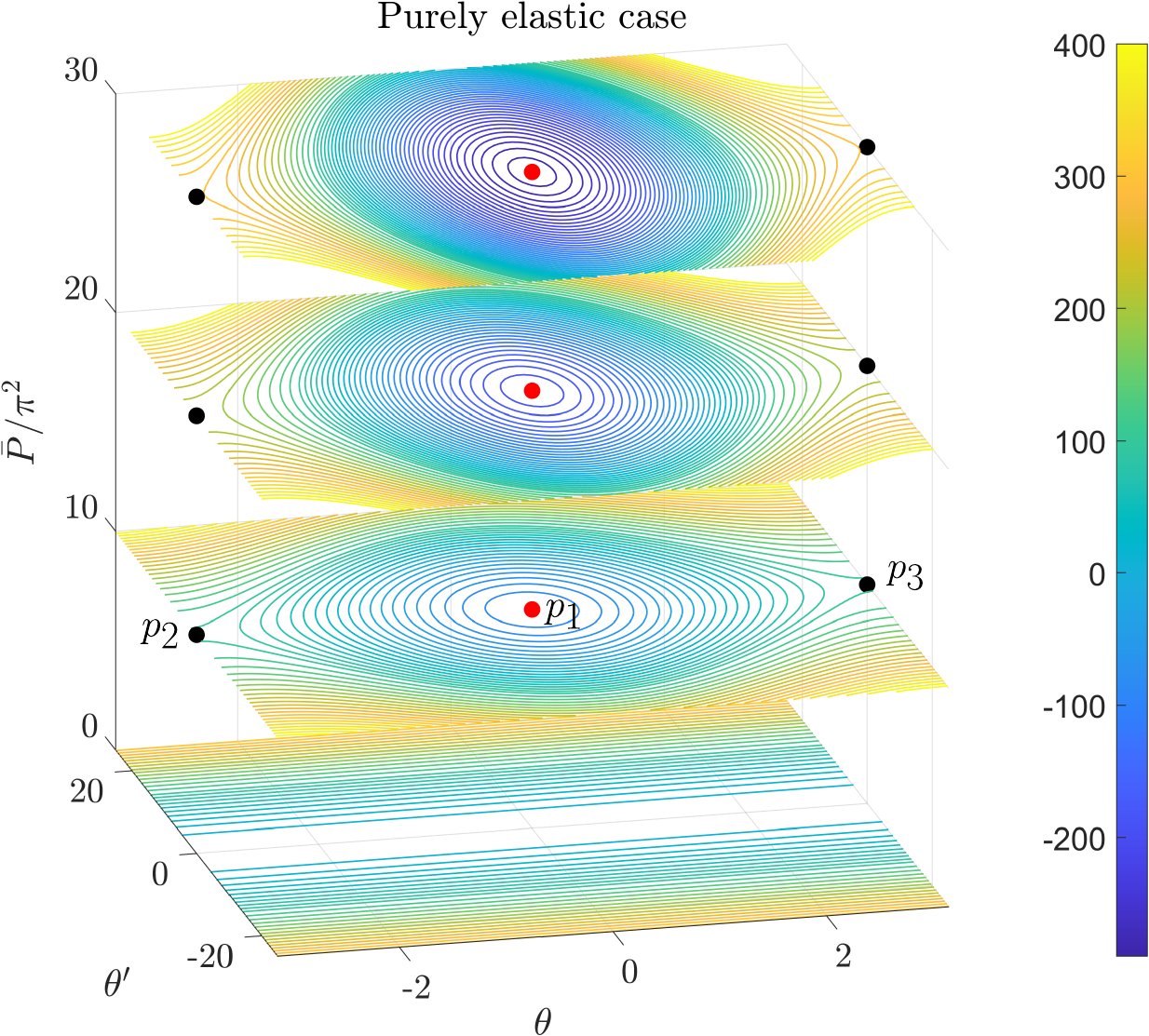}
	\caption{Snapshots of phase portrait for a purely elastic rod as $\bar{P}$ is varied. Red dot indicates center while black dot indicate saddle point.}
	\label{fig:section-view-elastic}
\end{figure}

We begin our analysis with the phase portrait corresponding to the purely elastic case ($\bar{K}_d = 0$). We compare our results with an earlier work done by Maddocks \cite{maddocks1981analysis}. For $\bar{P} > 0$, the critical points for the purely elastic case are $p_1$, $p_2$, and $p_3$, where $p_1$ is a center, while $p_2$ and $p_3$ are saddle points (see Fig. \ref{fig:section-view-elastic}). An individual case of a separatrix is highlighted in the phase portrait for the loading parameter $\bar{P} = \frac{\pi^2}{3}$ in Fig. \ref{fig:free-standing-elastica}. The separatrix passing through $p_2$ and $p_3$ divides the phase plane into two regions. In the first region inside the separatrix, the trajectories form closed orbits that enclose the origin $p_1$ and intersect the $\theta' = 0$ axis leading to inflectional elastica. For any trajectory with an inflection point, there exists a bound on $\theta$ such that $|\theta| < \pi$. In the second region outside the separatrix, the trajectories are unbounded and do not cross the $\theta' = 0$ axis. The resulting deformed configurations on such trajectories have no inflectional points. The heteroclinic orbits joining the saddle points ($p_2$ and $p_3$) coinciding with the separatrix produces shapes with localized deformations. Note that for a dynamical system, heteroclinic orbit represents a non-periodic trajectory.

At $\bar{P}= 0$, the phase portrait is depicted by a family of constant-$\theta'$ lines. All the points on $\theta'=0$ line are critical points. For $\bar{P} > 0$ (compressive), we have three critical points: the center $p_1$ and the saddle points $p_2$ and $p_3$. As $\bar{P}$ increases, the height of the region bounded by the separatrix containing $p_1$, also known as the `eye', increases along the $\theta'$-axis (see Fig. \ref{fig:section-view-elastic}). We now analyze the phase portrait of transverse $\vb*{h}_e$.

\subsection{Transverse $\vb*{h}_e$}
\begin{figure}[h!]
	\centering
	\includegraphics[width=0.9\linewidth]{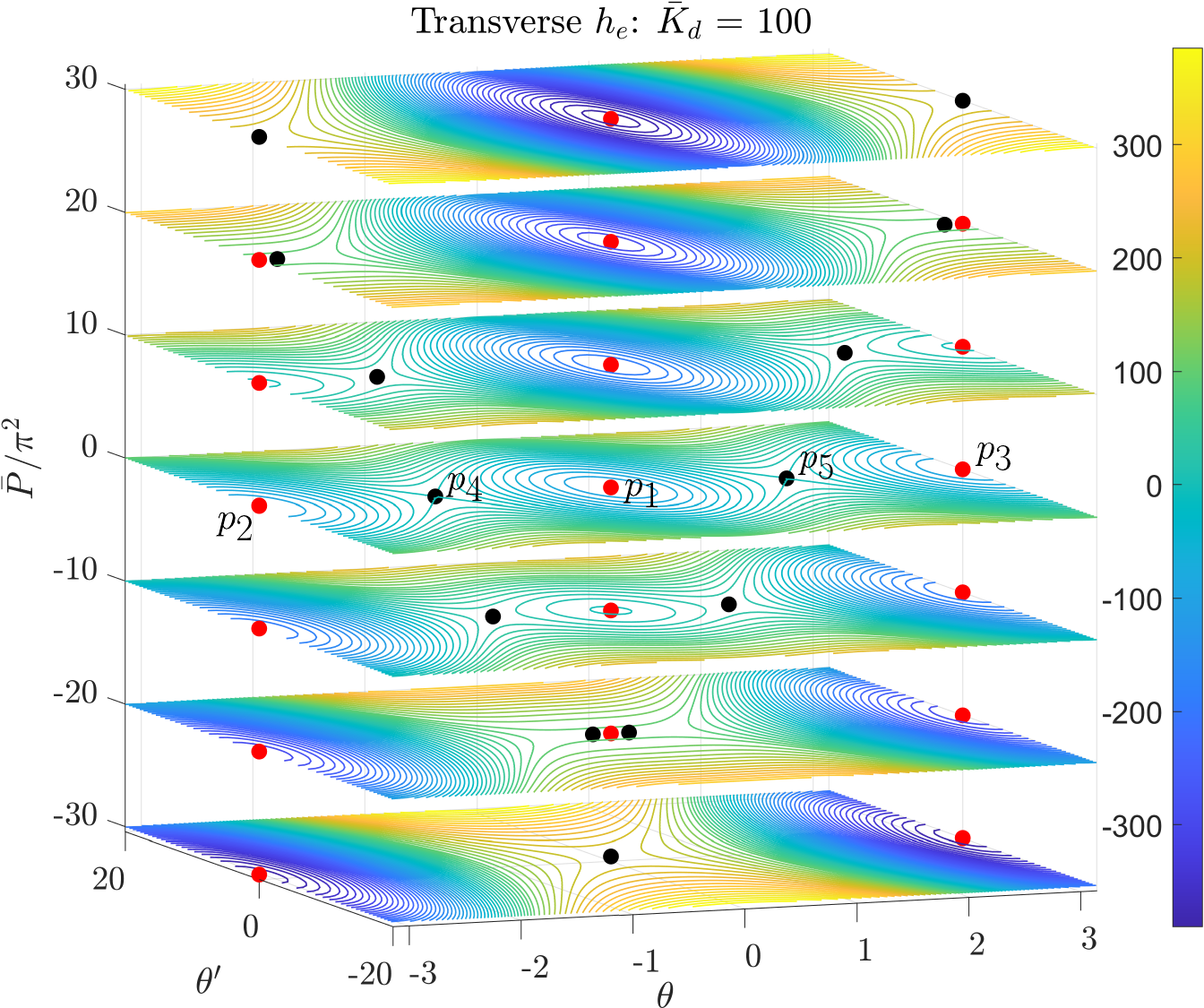}
	\caption{Phase portrait snapshots for a soft ferromagnetic rod in the presence of transverse $\vb*{h}_e$ as $\bar{P}$ is varied; $\bar{P}>0$ (compression) and $\bar{P}<0$ (tension).}
	\label{fig:section-view-transverse-kdbar-100}
\end{figure}

Fig. \ref{fig:section-view-transverse-kdbar-100} represents the phase portraits of a ferromagnetic rod subjected to transverse $\vb*{h}_e$ at various values of $\bar{P}$, ranging from $30\pi^2$ (compressive) to $-30\pi^2$ (tensile). Our analysis is conducted with $\bar{K}_d = 100$. At this value, it has been shown that the magnetic energy significantly competes with the elastic energy \cite{AvatarDabade2024}.

Initially, at $\bar{P} = 30\pi^2$ (compressive), there are three critical points: one center $p_1$, and two saddle points $p_2$ and $p_3$ (see Fig. \ref{fig:section-view-transverse-kdbar-100}). As compressive load is gradually reduced, we observe that the points $p_2$ and $p_3$ bifurcate at $\bar{P}= 20.26\pi^2$,  resulting in the emergence of two new critical points $p_4$ and $p_5$. This value is derived from the expression $\bar{P} - 2\bar{K}_d = 0$. After the bifurcation, $p_2$ and $p_3$ become centers while $p_4$ and $p_5$ are saddle points. As $\bar{P}$ continues to decrease, $p_4$ and $p_5$ drift towards $p_1$. They eventually merge with $p_1$ at $\bar{P} = -20.26\pi^2$, which is obtained by solving the equation $\bar{P} + 2\bar{K}_d = 0$. Consequently, point $p_1$ transitions into a saddle point.

This leaves us with three critical points that are identical to those in the purely elastic case. Therefore, as $\bar{P}$ is decreased from $30\pi^2$ to $-30\pi^2$, this indicates a subcritical pitchfork bifurcation. This resembles a second-order phase transition with $\bar{P}$ as the order parameter. Hence, the magnetization alters the phase portrait across a range of values for $\bar{P}$, with its extent governed by the parameter $\bar{K}_d$. We now proceed to the study of phase portrait of the longitudinal $\vb*{h}_e$ case.

\subsection{Longitudinal $\vb*{h}_e$}
\begin{figure}[h!]
	\centering
	\includegraphics[width=0.9\linewidth]{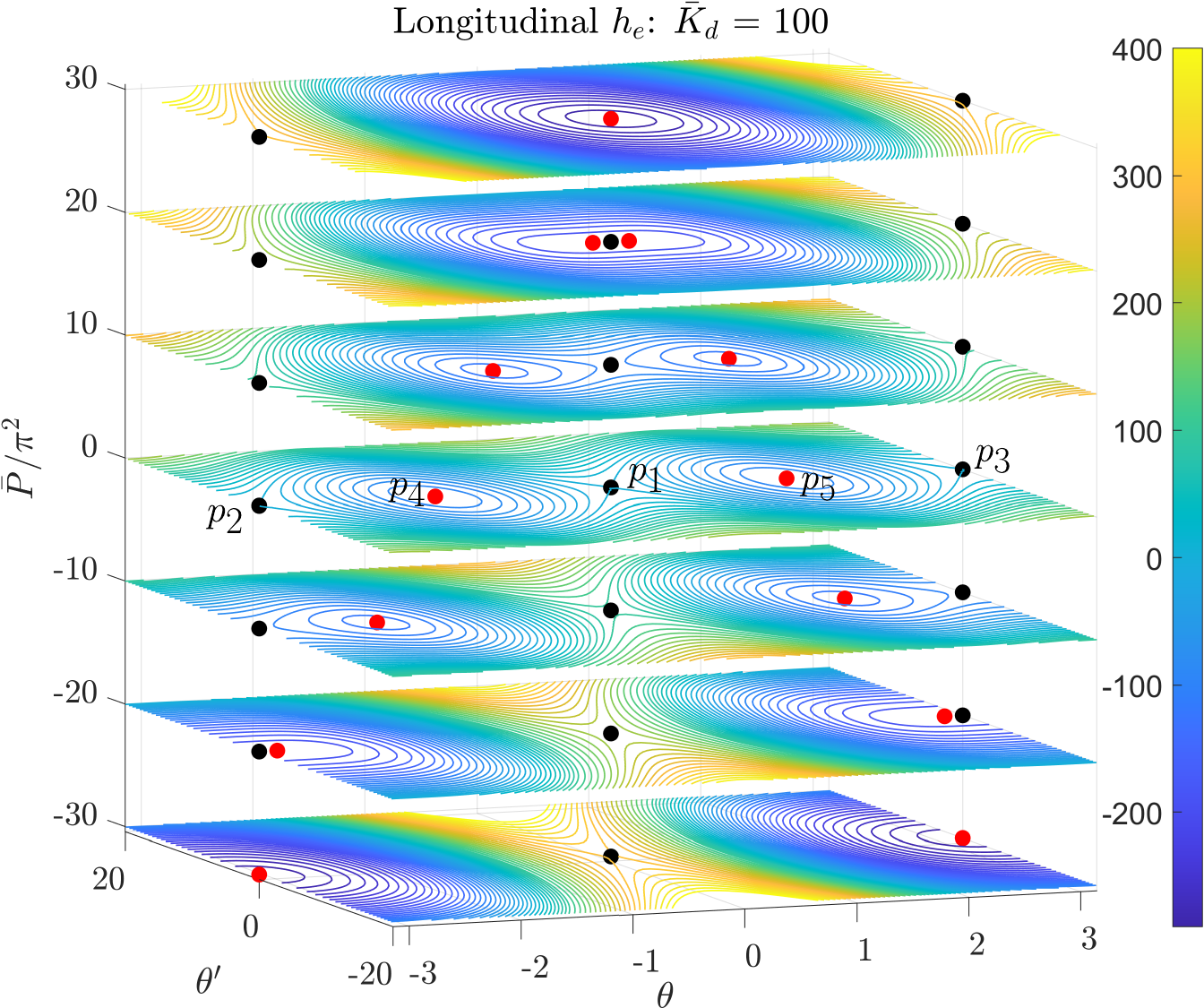}
	\caption{Phase portrait snapshots of a soft ferromagnetic rod subjected to longitudinal $\vb*{h}_e$ as $\bar{P}$ is varied; $\bar{P}>0$ (compression) and $\bar{P}<0$ (tension).}
	\label{fig:section-view-longitudinal-kdbar-100}
\end{figure}

 Fig. \ref{fig:section-view-longitudinal-kdbar-100} depicts the phase portraits of a ferromagnetic rod subjected to longitudinal $\vb*{h}_e$ as $\bar{P}$ varies from $30\pi^2$ to $-30\pi^2$. We fix the value of $\bar{K}_d = 100$ for our analysis.

Initially, at $\bar{P} = 30\pi^2$, there are three critical points: one center $p_1$, and two saddle points $p_2$ and $p_3$.  As $\bar{P}$ decreases, the saddle point $p_1$ bifurcates to form two new critical points ($p_4$ and $p_5$) at $\bar{P} = -20.26\pi^2$. This behaviour contrasts with the transverse $\vb*{h}_e$ case. Following the bifurcation, $p_1$ transitions to a saddle point, while $p_4$ and $p_5$ become centers. As $\bar{P}$ continues to decrease, $p_4$ and $p_5$ drift towards $p_2$ and $p_3$, respectively, eventually merging with them at $\bar{P} = -20.26\pi^2$. The points $p_2$ and $p_3$ become centers. Ultimately, we are left with three critical points.

Therefore, as $\bar{P}$ is decreased from $30\pi^2$ to $-30\pi^2$, this indicates a supercritical pitchfork bifurcation in the longitudinal $\vb*{h}_e$ case. Similar to the transverse $\vb*{h}_e$ case, this too exemplifies a second-order phase transition governed by the order parameter $\bar{P}$.

\section{Free-standing elastica}
We begin this section by describing the methodology we adopted to obtain the free-standing deformed configurations of the planar ferromagnetic rod using the trajectories in the Hamiltonian phase portrait. We study the following cases: (a) a purely elastic rod, (b) a ferromagnetic rod subjected to transverse $\vb*{h}_e$, and (c) a ferromagnetic rod exposed to longitudinal $\vb*{h}_e$. We keep the parameters $\bar{P}$ and $\bar{K}_d$ fixed for our analysis, and set their values to $\bar{P} = \frac{\pi^2}{3}$ and $\bar{K}_d = 100$.

To obtain the free-standing shapes, we first choose a constant-Hamiltonian curve in the phase portrait corresponding to purely elastic, transverse $\vb*{h}_e$ and longitudinal $\vb*{h}_e$ cases, see Figs. \ref{fig:free-standing-elastica}, \ref{fig:free-standing-transverse-kdbar-100} and \ref{fig:free-standing-longitudinal-kdbar-100} respectively. We then integrate along the curve for the desired length of the trajectory on the phase portrait to compute the free-standing shapes. For a purely elastic rod ($\bar{K}_d = 0$), the trajectory is obtained from Eqn. \ref{eqn:transverse-hamiltonian} for some chosen value $C = k$ as follows:
\begin{equation}
	\begin{split}
		\dv{\theta}{\bar{s}} &= \sqrt{2}\sqrt{\bar{P}\cos\theta(\bar{s}) + k}, \\
	\implies	\int_{0}^{\bar{s}}\text{d}\bar{s} &= \int_{\theta_0}^{\theta} \frac{\text{d}\theta}{\sqrt{2}\sqrt{\bar{P}\cos\theta(\bar{s}) + k}},
	\end{split} \label{eqn:arclength-theta-equation-elastic}
\end{equation}
where $\theta_0 = \theta(\bar{s}=0)$ is determined from the initial conditions.

For a ferromagnetic elastic rod exposed to transverse $\vb*{h}_e$, the trajectory is obtained from Eqn. \ref{eqn:transverse-hamiltonian} for $C = k$ as follows:
\begin{equation}
	\begin{split}
		\dv{\theta}{\bar{s}} &= \sqrt{2}\sqrt{\bar{K}_d \cos^2\theta(\bar{s}) + \bar{P}\cos\theta(\bar{s}) + k}, \\
	\implies	\int_{0}^{\bar{s}}\text{d}\bar{s} &= \int_{\theta_0}^{\theta} \frac{\text{d}\theta}{\sqrt{2}\sqrt{\bar{K}_d \cos^2\theta(\bar{s}) + \bar{P}\cos\theta(\bar{s}) + k}}. \label{eqn:arclength-theta-equation-transverse}
	\end{split}
\end{equation}
Similarly, the trajectory corresponding to longitudinal $\vb*{h}_e$ is derived from Eqn. \ref{eqn:longitudinal-hamiltonian} as
\begin{equation}			\int_{0}^{\bar{s}}\text{d}\bar{s} = \int_{\theta_0}^{\theta} \frac{\text{d}\theta}{\sqrt{2}\sqrt{\bar{K}_d \sin^2\theta(\bar{s}) + \bar{P}\cos\theta(\bar{s}) + k}}.  \label{eqn:arclength-theta-equation-longitudinal}
\end{equation}
The Cartesian coordinates of the rod centerline are computed as follows:
\begin{equation}
	z(\bar{s}) = \int_{0}^{\bar{s}}\cos\theta(\bar{s}')d\bar{s}', \quad y(\bar{s}) = \int_{0}^{\bar{s}}\sin\theta(\bar{s}')d\bar{s}'. \label{eqn:centerline-coordinates}
\end{equation}
In all the Eqns. \ref{eqn:arclength-theta-equation-elastic}, \ref{eqn:arclength-theta-equation-transverse} and \ref{eqn:arclength-theta-equation-longitudinal}, $\theta$ is the independent variable, while $\bar{s}$ is the dependent variable. Therefore, we must first determine the minimal interval of integration for the independent variable $\theta$, specifically, $\theta \in [\theta_0, \theta_1]$, which is contained within the chosen constant-Hamiltonian curve $\mathcal{H}(\theta, \theta') = k$. The trajectories can be classified into the following types:
\begin{enumerate}[label={(\arabic*)}]\label{types}
    \item Trajectories with inflection points: These are trajectories encircling the centers. For these trajectories, we identify $\theta_0$ and $\theta_1$ as the points of intersection of the $\mathcal{H}(\theta, \theta') = k$ curve on the $\theta$-axis. We exploit the symmetric nature of the phase portrait to construct an extended interval of integration, $\theta \in [\theta_{\text{initial}}, \theta_{\text{final}}]$, where $\theta_{\text{initial}}$ and $\theta_{\text{final}}$ denote the initial and the final points of the trajectory chosen for integration.
    \item Trajectories without inflection points: In this case, we can set $\theta_0 = 0$ and let $\theta_1$ be any value greater than $\theta_0$ that is required for integrating along the desired length of trajectory. Here, we have $\theta_{\text{initial}} = \theta_0$ and $\theta_{\text{final}} = \theta_1$.
    \item Homoclinic and heteroclinic orbits: These are non-periodic orbits. For these orbits, we follow a similar approach as in scenario (1) to determine $\theta_0$ and $\theta_1$ for the interval $[\theta_0, \theta_1]$. We then reverse this interval and append it to $[\theta_0, \theta_1]$ to create an extended interval, denoted as $[\theta_{\text{initial}}, \theta_{\text{final}}]$. When these trajectories approach a saddle point $(\theta,\theta') = (\pm\theta_{\text{saddle}},0)$, the arc length parameter extends to infinity:
    \begin{equation}
        \theta \rightarrow \pm\theta_{\text{saddle}} \implies \bar{s} \implies \mp \infty.
    \end{equation}
The deformed configurations corresponding to these trajectories represent infinitely long rods subjected solely to terminal forces.
\end{enumerate}

When the chosen trajectory is a separatrix, numerical singularity issues occur at the points of intersection on the $\theta$-axis due to the separatrix since the denominator of the integrands in Eqns. \ref{eqn:arclength-theta-equation-elastic},\ref{eqn:arclength-theta-equation-transverse} and \ref{eqn:arclength-theta-equation-longitudinal} evaluate to zero. To circumvent this problem, we offset the limit points of integration, $\theta_0$ and $\theta_1$, slightly from the actual intersection points by approximately $10^{-8}$.

We then numerically solve the ordinary differential equations (Eqns. \ref{eqn:arclength-theta-equation-elastic}, \ref{eqn:arclength-theta-equation-transverse} and \ref{eqn:arclength-theta-equation-longitudinal}) using MATLAB's built-in \texttt{ode89} function. The \texttt{ode89} algorithm \cite{Verner2009} is a 9th-order accurate scheme and is highly efficient for integrating over long intervals or when tight tolerances are required. The numerical solution for $\bar{s}$ is obtained at discrete values of $\theta$ within the interval $[\theta_{\text{initial}}, \theta_{\text{final}}]$. This discrete solution is represented as $\{(s_{(i)}, \theta_{(i)})\}$, where $i = 1, 2, \dots, N$ such that $s_{(0)} = 0$, $\theta_{(0)} = \theta_{\text{initial}}$, and $\theta_{(N)} = \theta_{\text{final}}$; $N$ is the number of discrete points. Finally, the coordinates of the rod centerline are obtained by numerically integrating Eqns. \ref{eqn:centerline-coordinates} using the standard trapezoidal method over the discrete solution set $\{(s_{(i)}, \theta_{(i)})\}_{i=1}^{N}$.

\subsection*{(a) Purely elastic case}
\begin{figure}[h!]
	\centering
	\includegraphics[width=\linewidth]{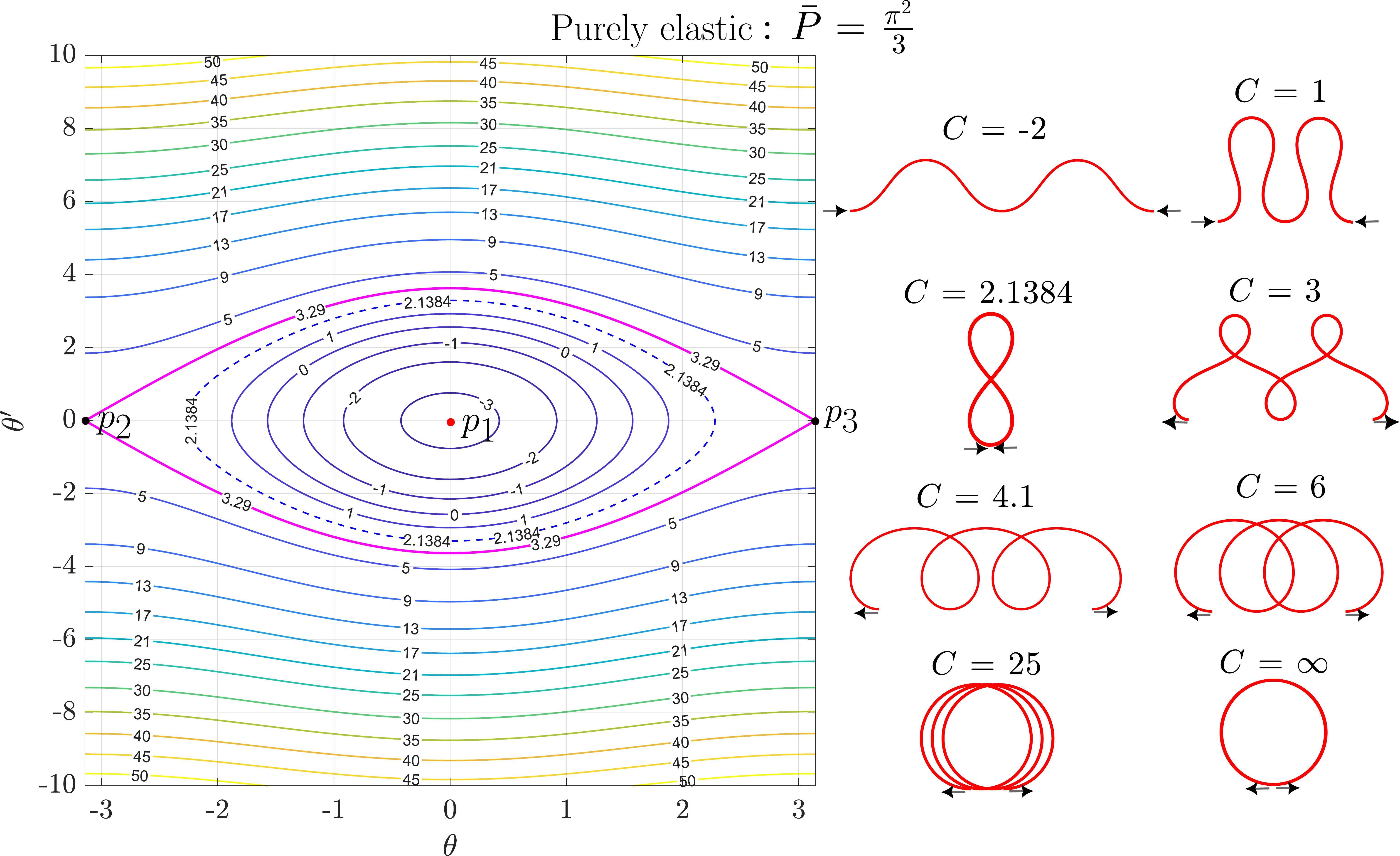}
	\caption{Purely elastic rod ($\bar{P} = \frac{\pi^2}{3}$): Phase portrait with the separatrix highlighted in magenta and free-standing deformed shapes. Arrows ($\rightarrow$) indicate the loading direction. $\bar{P}$ is compressive for trajectories inside the dashed closed orbit and tensile for those outside.}
	\label{fig:free-standing-elastica}
\end{figure}

In Fig. \ref{fig:free-standing-elastica}, we present the Hamiltonian phase portrait and the free-standing deformed shapes for a purely elastic rod. The type (1) trajectories within the separatrix, highlighted in magenta, generate shapes with inflection points. We observe self-intersections during deformations, which have been replicated in meticulously designed experiments on planar elastic deformation of ribbons by Bigoni et al. \cite{Bigoni2015}.

The center, $p_1$, represents an undeformed (trivial) configuration. As the value of $C$ increases from this trivial scenario —- where $C$ is at its minimum -— deformed configurations begin to emerge. For trajectories in the immediate neighbourhood of center ($p_1$), the free-standing shapes experience a compressive load ($\bar{P}>0$). As $C$ increases from its value at the origin, the ends of the shapes draw closer together. On the trajectory with $C = 2.1384$, represented by the dashed closed orbit in Fig. \ref{fig:free-standing-elastica}), the nature of loading $\bar{P}$ transitions from compressive to tensile ($\bar{P} < 0$) for the buckled configurations.  This transition causes the ends of the shapes to meet. When $C > 2.1384$, the left and right ends of the undeformed rod exchange their relative positions. This results in a tensile state for the rod. The type (2) trajectories result in deformations which are circular in nature. The free-standing shapes for the purely elastic case obtained in this study are consistent with those reported in the literature \cite{Harmeet2018}.

\begin{figure}[h!]
	\centering
	\includegraphics[width=\linewidth]{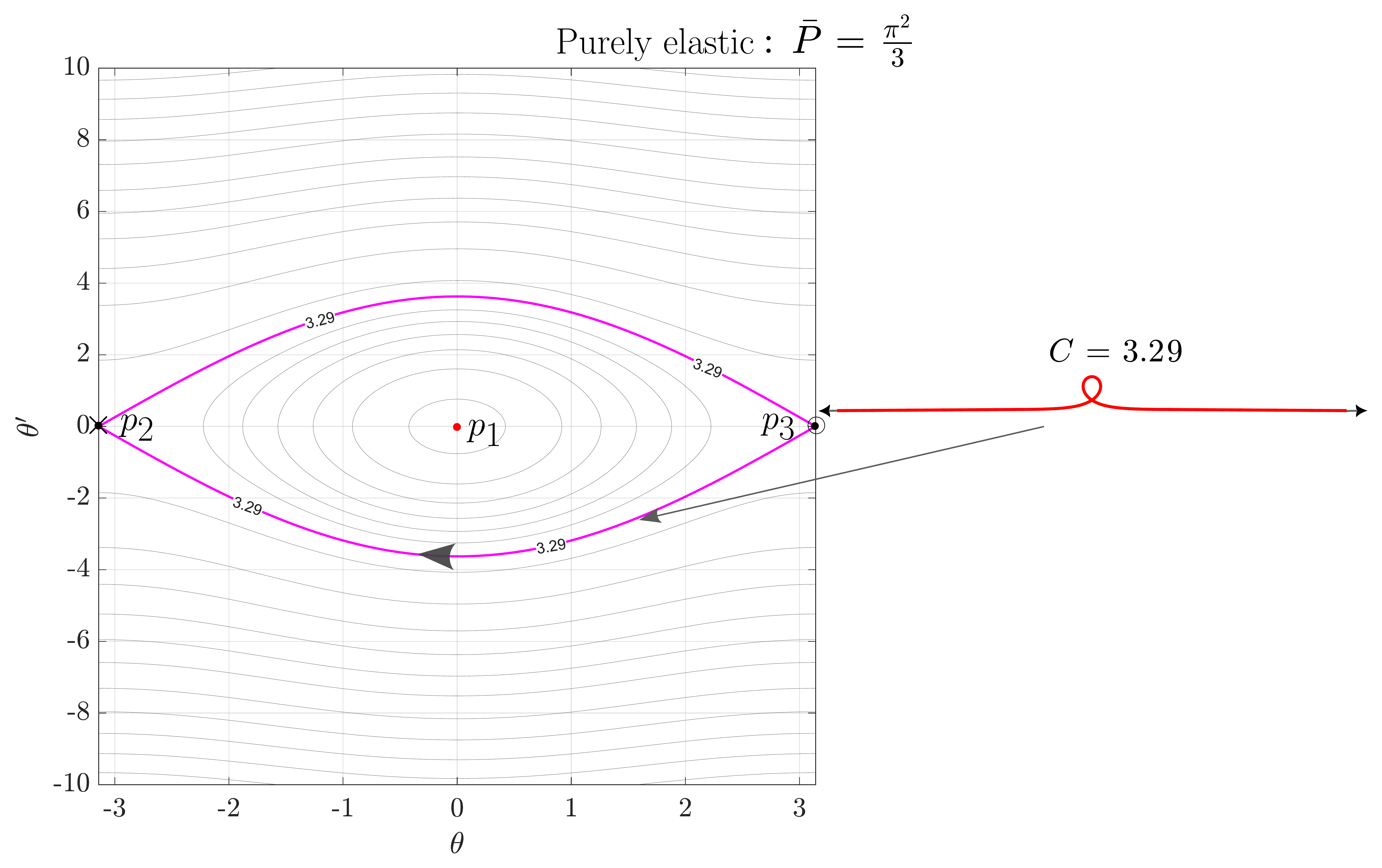}
	\caption{Purely elastic case: Localised deformed shape corresponding to a heteroclinic orbit coincinding with the separatrix at $C=3.29$.}
	\label{fig:localised-elastic}
\end{figure}

The type (3) trajectory that coincides with the separatrix at $C = 3.29$ results in a localized buckled shape, as shown in Fig. \ref{fig:localised-elastic}. This localized shape was obtained by integrating anticlockwise along the heteroclinic orbit located on the lower portion of the separatrix, which connects the saddle points, $p_2$ and $p_3$. The choice of integration path influences the deformed shape only by producing a mirror image about the $\theta$-axis. Since the heteroclinic orbit connects the saddle points ($p_2$ and $p_3$), $(\theta, \theta') = (\pm \pi, 0)$, the arc length parameter extends to infinity:
\begin{equation}
	\theta \rightarrow \pm \pi \implies \bar{s} \rightarrow \mp \infty.
\end{equation}
Next, we discuss the free-standing deformations of a ferromagnetic rod in the presence of transverse $\vb*{h}_e$.

\subsection*{(b) Transverse $\vb*{h}_e$ case}
\begin{figure}[h!]
	\centering
	\includegraphics[width=\linewidth]{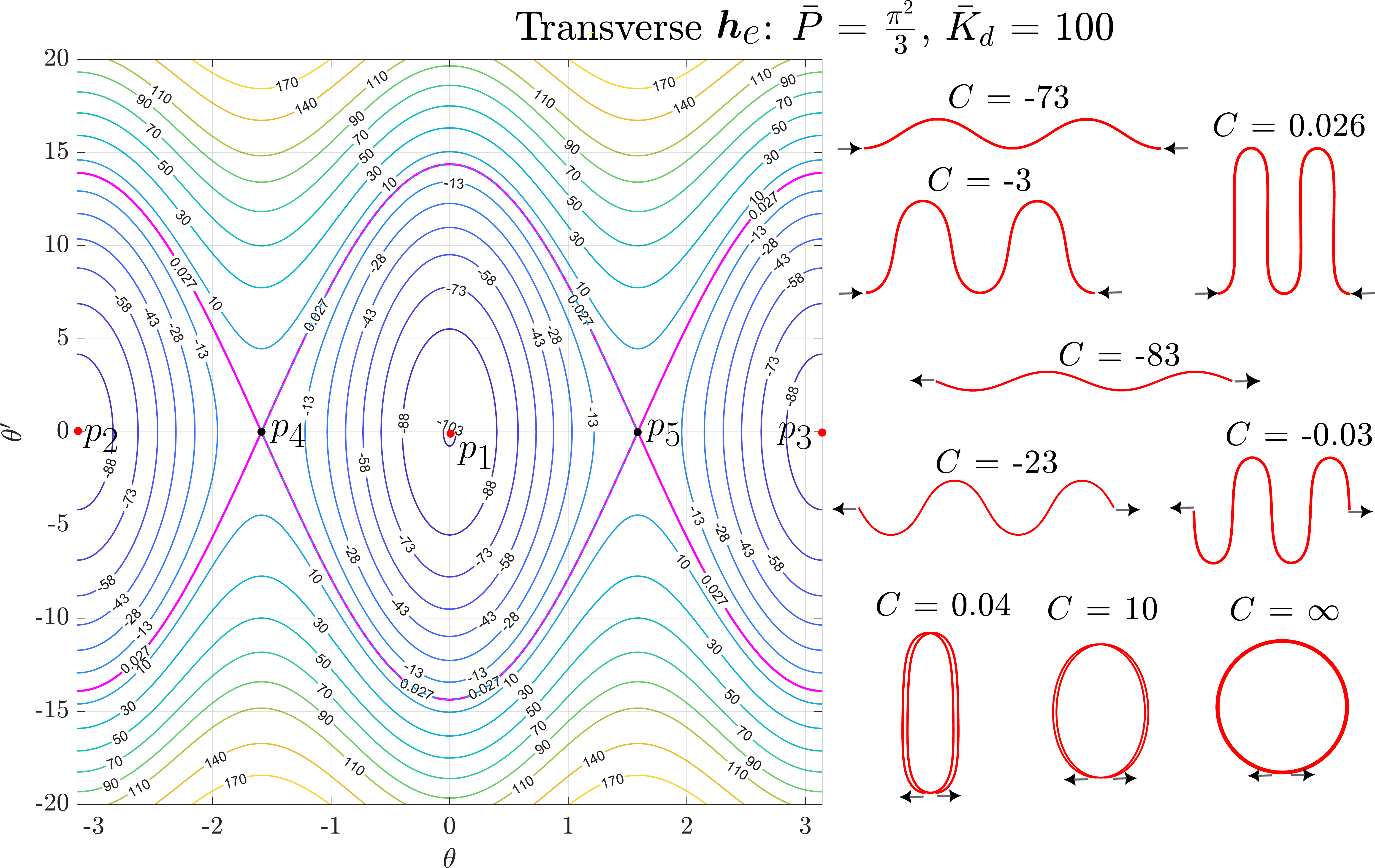}
	\caption{Transverse $\vb*{h}_e$ ($\bar{P} = \frac{\pi^2}{3}$, $\bar{K}_d = 100$): Phase portrait with separatrices highlighted in magenta, and free-standing deformed shapes. $\bar{P}$ is compressive for trajectories inside the dashed orbit (coinciding with the separatrix enclosing $p_1$) and is tensile for those outside.}
	\label{fig:free-standing-transverse-kdbar-100}
\end{figure}

Fig. \ref{fig:free-standing-transverse-kdbar-100} presents the Hamiltonian phase portrait of a ferromagnetic rod subjected to transverse $\vb*{h}_e$ for $\bar{P} = \frac{\pi^2}{3}$ and $\bar{K}_d = 100$. Comparing with the purely elastic case, we have three centers ($p_1$, $p_2$ and $p_3$) and two saddle points ($p_4$ and $p_5$) in the transverse $\vb*{h}_e$ case. The separatrix is highlighted in magenta in Fig. \ref{fig:free-standing-transverse-kdbar-100}.

We illustrate the free-standing shapes of the deformed rod corresponding to the different types of trajectories in Fig. \ref{fig:free-standing-transverse-kdbar-100}. Trajectories of type (1) exhibit deformations with inflection points and are subjected to compressive loading ($\bar{P} > 0$). As $C$ increases from its value at the origin ($p_1$), the ends of the deformed rod approach each other. On the separatrix enclosing $p_1$, the nature of loading switches from compressive ($\bar{P}>0$) to tensile ($\bar{P}<0$), as indicated by the dashed orbit in Fig. \ref{fig:free-standing-transverse-kdbar-100}. Note that the separatrix coincides with the dashed orbit. For type (1) trajectories with centers at $p_2$ and $p_3$, the deformed rod configurations also display inflection points. In this case, the terminal ends of the rod cross each other, resulting in a tensile loading condition. The shapes resulting from type (2) shapes lack inflection points. As the value of $C$ increases, these deformations transition from oval-like to a circular shape as $C\rightarrow \infty$.

\begin{figure}[h!]
	\centering
	\includegraphics[width=1\linewidth]{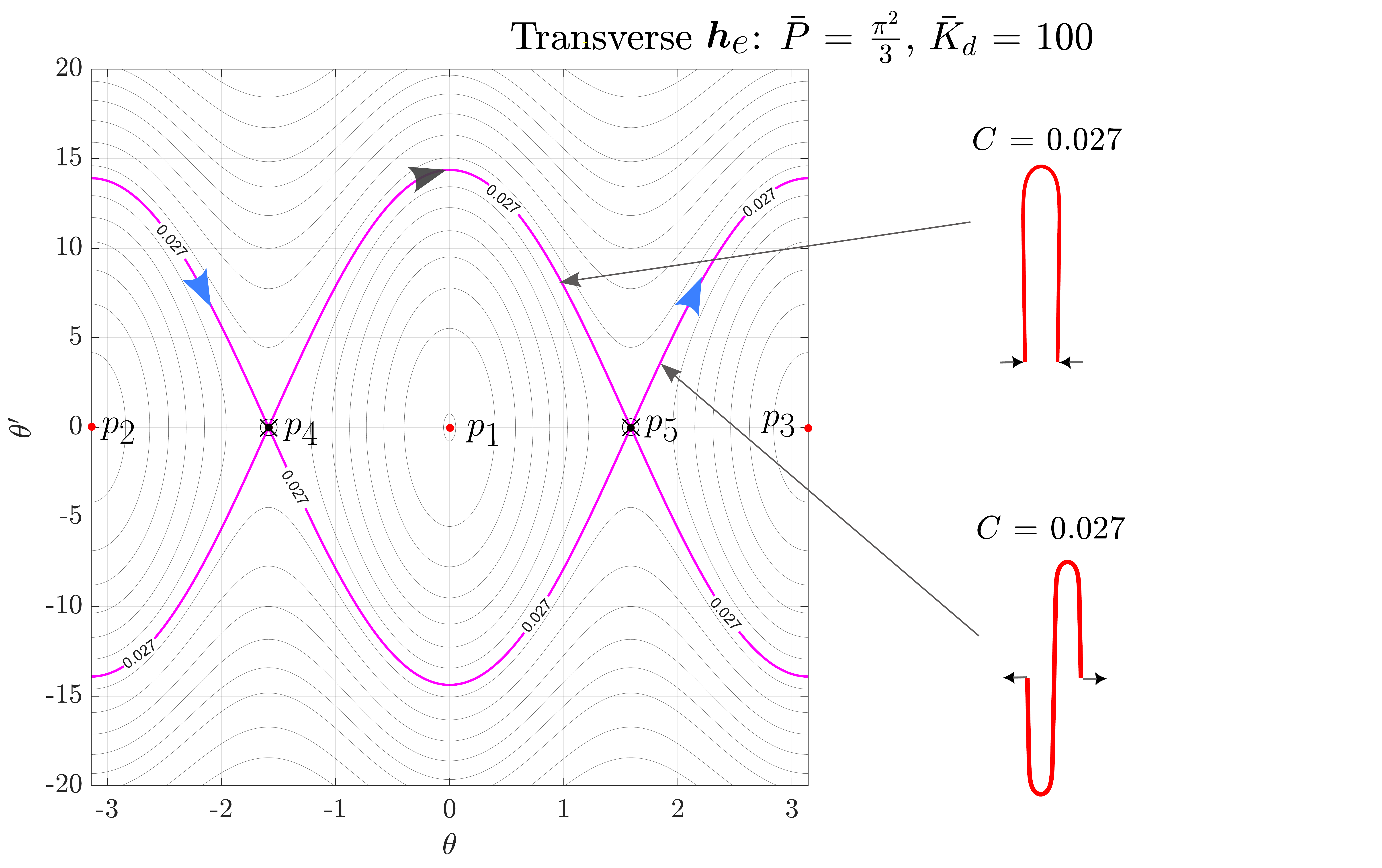}
	\caption{Transverse $\vb*{h}_e$ ($\bar{P} = \frac{\pi^2}{3}$, $\bar{K}_d = 100$): Localized shapes obtained by integrating along the trajectory on the separatrix enclosing $p_1$ (indicated by a black arrowhead) and the on e enclosing $p_2$ and $p_3$ (indicated by a blue arrowhead).}
	\label{fig:localised-transverse-kdbar-100}
\end{figure}

The localized shapes corresponding to the heteroclinic orbits on the separatrices, that is, type (3) trajectories, are shown in Fig. \ref{fig:localised-transverse-kdbar-100}. Both shapes exhibit steep deflections. For the shape produced by the trajectory coinciding with the separatrix enclosing the origin, the loading $\bar{P}$ is compressive in nature, and the shape is symmetric. In contrast, for the shape corresponding to the separatrix enclosing $(\pm \pi, 0)$, the loading $\bar{P}$ is tensile, and the corresponding shape is anti-symmetric. Finally, we discuss about the ferromagnetic rod subjected to longitudinal $\vb*{h}_e$ field.

\subsection*{(c) Longitudinal $\vb*{h}_e$ case}
\begin{figure}[h!]
	\centering
	\includegraphics[width=\linewidth]{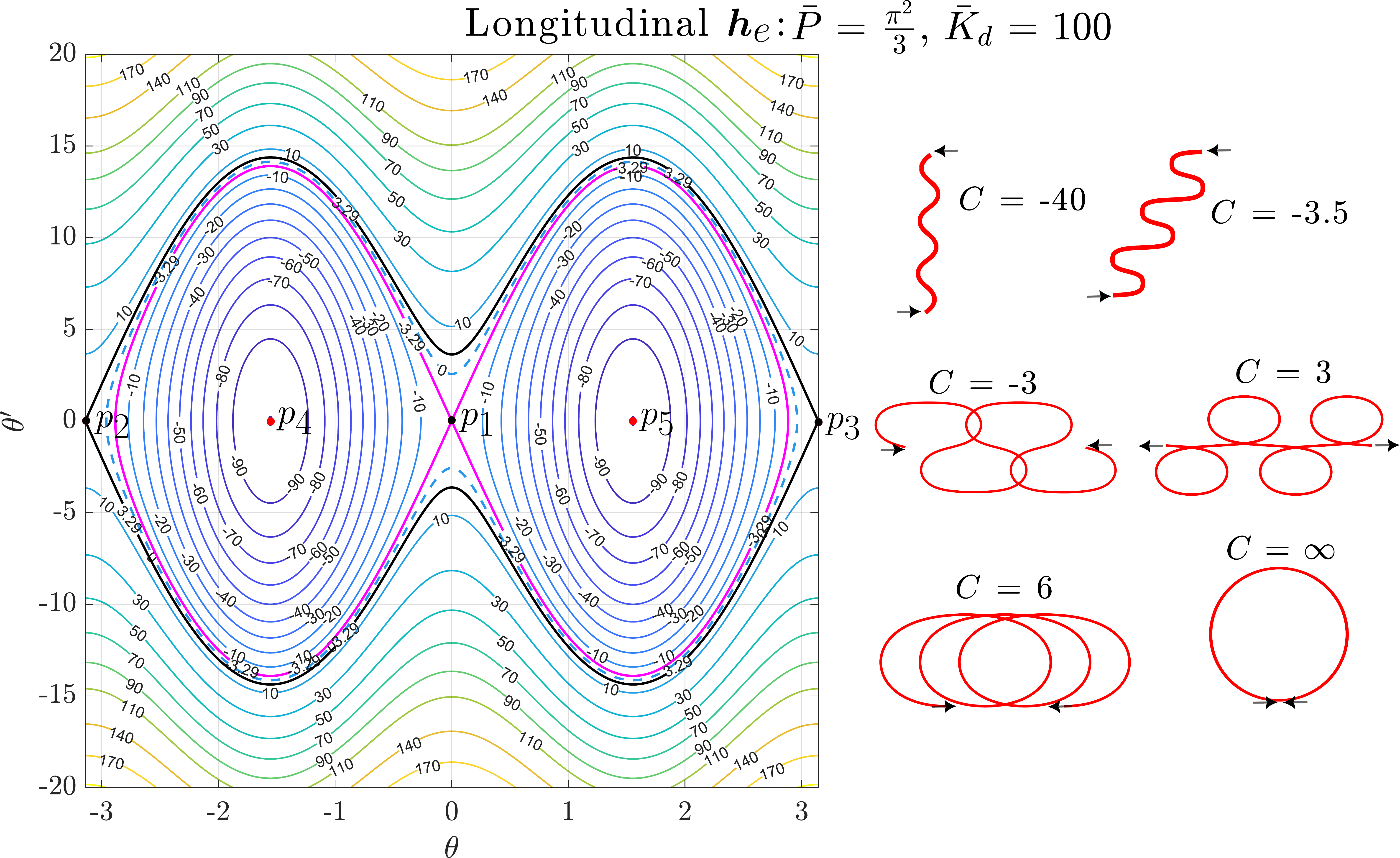}
	\caption{Longitudinal $\vb*{h}_e$ ($\bar{P} = \frac{\pi^2}{3}$, $\bar{K}_d = 100$): Phase portrait with separatrices coloured in magenta and black, and free-standing deformed shapes.}
	\label{fig:free-standing-longitudinal-kdbar-100}
\end{figure}

Fig. \ref{fig:free-standing-longitudinal-kdbar-100} depicts the phase portrait of a ferromagnetic rod subjected to a longitudinal $\vb*{h}_e$ for $\bar{P} = \frac{\pi^2}{3}$ and $\bar{K}_d = 100$. The phase portrait features two centers, denoted as $p_4$ and $p_5$, along with three saddle points identified as $p_1$, $p_2$ and $p_3$. In contrast to the purely elastic and transverse $\vb*{h}_e$ cases, this phase portrait reveals the presence of two separatrices. The two separatices -- inner and outer -- are highlighted in magenta and black respectively.

The free-standing deformed shapes of the rod are also shown in Fig. \ref{fig:free-standing-longitudinal-kdbar-100}. The deformations due to type (1) trajectories with centers ($p_4$ and $p_5$) are helical and possess inflection points. With increase in $C$ from its value at $p_4$ and $p_5$, the deformed shapes begin to tilt sideways. In contrast, type (1) trajectories lying between the two separatrices undergo self-intersections. As $C$ increases from its value on the inner separatrix, the terminal ends of the buckled shapes approach each other. At $C = 0$, the ends of the rod meet, and the nature of loading $\bar{P}$ switches from compressive to tensile. For shapes with $C > 0$, the left and right ends switch their relative positions. As mentioned earlier, the deformations generated from type (2) trajectories outside the outer separatrix have no inflection points. As $C \rightarrow \infty$, these shapes tend to form a circular ring.

\begin{figure}[h!]
	\centering
	\includegraphics[width=1\linewidth]{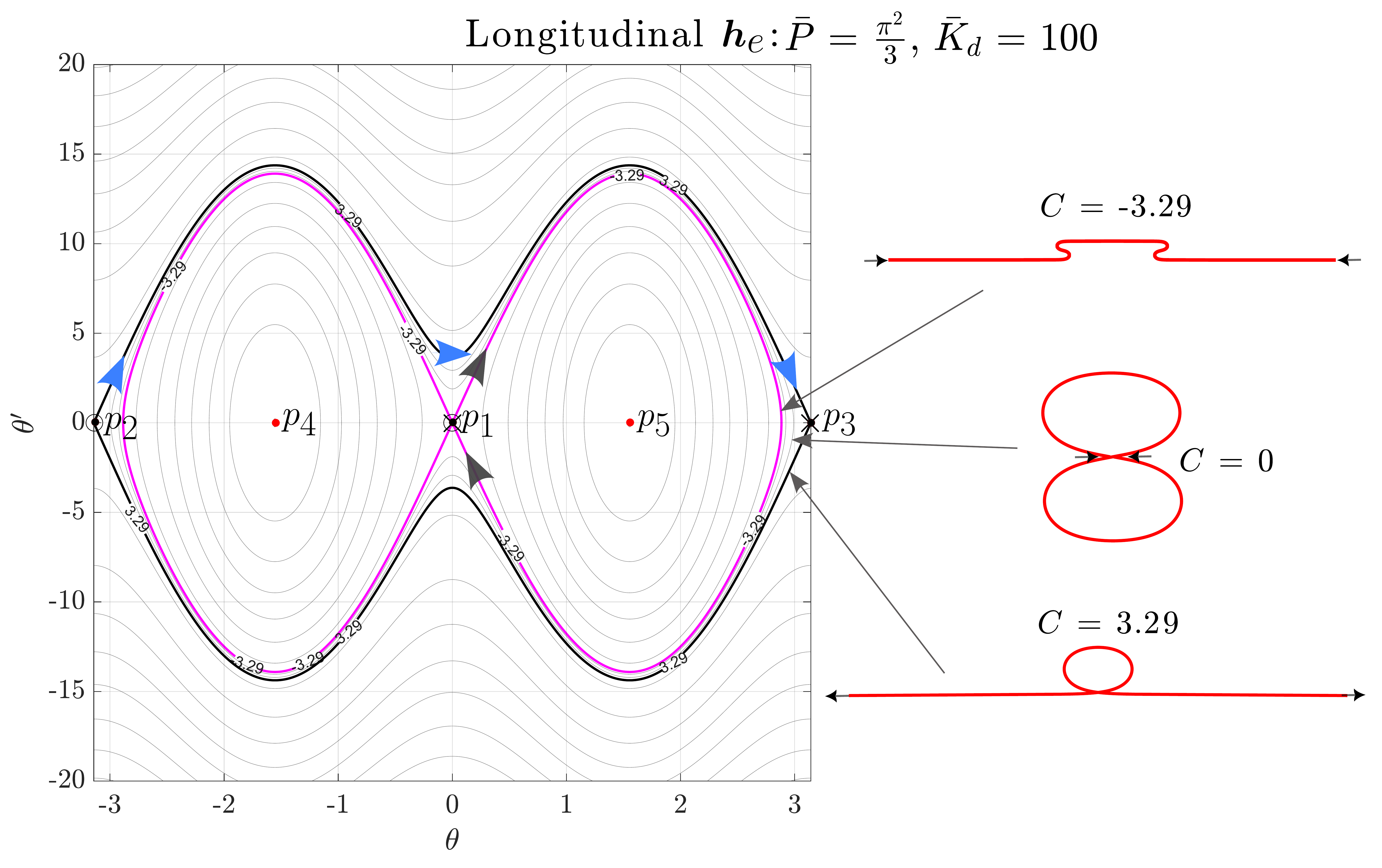}
	\caption{Longitudinal $\vb*{h}_e$ ($\bar{P} = \frac{\pi^2}{3}$, $\bar{K}_d = 100$): Localised shapes obtained by integrating along the homoclinic orbit coinciding with the inner separatrix (highlighted in magenta)  and the heteroclinic orbit (shown in black). A shape determined by integrating along a trajectory $C=0$ within the inner and outer separatrices is shown.}
	\label{fig:localised-longitudinal-kdbar-100}
\end{figure}

The type (3) trajectories lying on the separatrices ($C = \pm 3.29$) exhibit localized buckled modes, as shown in Fig. \ref{fig:localised-longitudinal-kdbar-100}. The localized shape for the inner separatrix is obtained by integrating clockwise along the homoclinic orbit at the saddle point $p_1$, where the arc length parameter $\bar{s}$ extends to infinity:
\begin{equation}
    \theta \rightarrow 0 \implies \bar{s} \rightarrow \pm \infty.
\end{equation}
Here, the loading parameter $\bar{P}$ is compressive in nature.  The shape corresponding to the outer separatrix is determined by integrating clockwise along the heteroclinic orbit that connects the saddle points ($p_2$ and $p_3$), with $\bar{P}$ being tensile in this case. Fig. \ref{fig:localised-longitudinal-kdbar-100} also shows a double-looped shape corresponding to a periodic trajectory of type (1) with its ends meeting.

\section{Boundary value problems}\label{sec:bvp}
In this section, we demonstrate how Hamiltonian phase portraits can be used to solve boundary value problems related to the planar deformation of ferromagnetic elastic rods. We focus on the deformation of planar ferromagnetic rods under three canonical boundary conditions. These conditions represent typical scenarios encountered in practice. The three canonical boundary conditions we will examine are:
\begin{enumerate}
	\item fixed-free or Euler's strut: $\theta(\bar{s} = 0) = 0, \theta'(\bar{s}=1) = 0$,
	\item pinned-pinned: $\theta'(\bar{s} = 0) = 0, \theta'(\bar{s}=1) = 0$,
	\item fixed-fixed: $\theta(\bar{s} = 0) = 0, \theta(\bar{s}=1) = 0$.
\end{enumerate}
As before, we consider three distinct loading scenarios: (i) purely elastic, (ii) transverse $\vb*{h}_e$, and (iii) longitudinal $\vb*{h}_e$. We determine the deformed configurations of the rod corresponding to the first mode for all these scenarios. We also ensure that the total non-dimensional arc length of the rod is maintained at unity.

\begin{figure}[h!]
	\centering
	\includegraphics[width=0.9\linewidth]{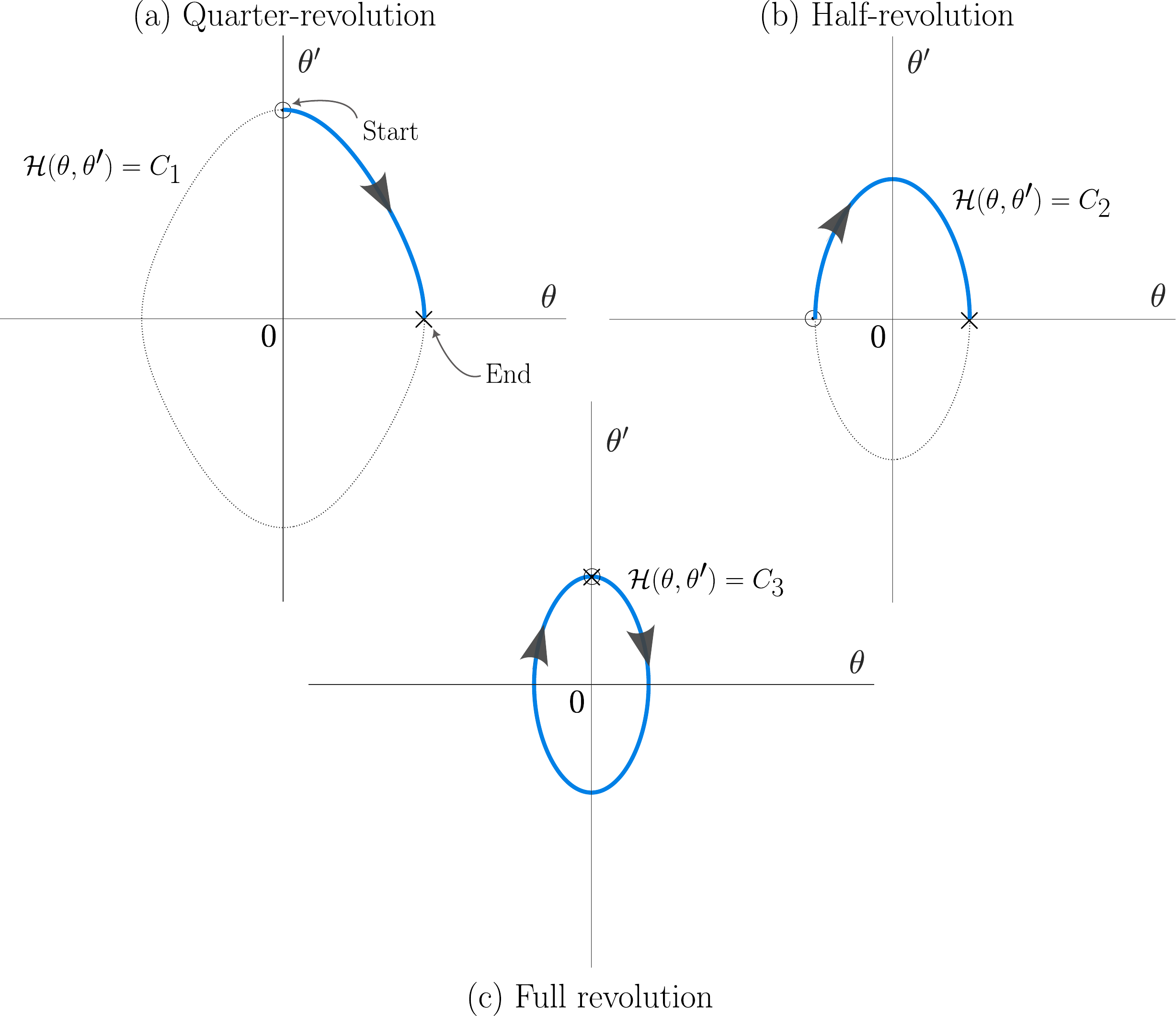}
	\caption{Schematic of trajectory integration showing (a) quarter-revolution, (b) half-revolution and (c) full revolution;  $C_1 > C_2 > C_3$. Note that for each of the aforementioned cases, the total arc length (highlighted in blue) of the deformed shape is 1.}
	\label{fig:schematic-trajectory}
\end{figure}

\begin{figure}[h!]
	\centering
	\includegraphics[width=0.9\linewidth]{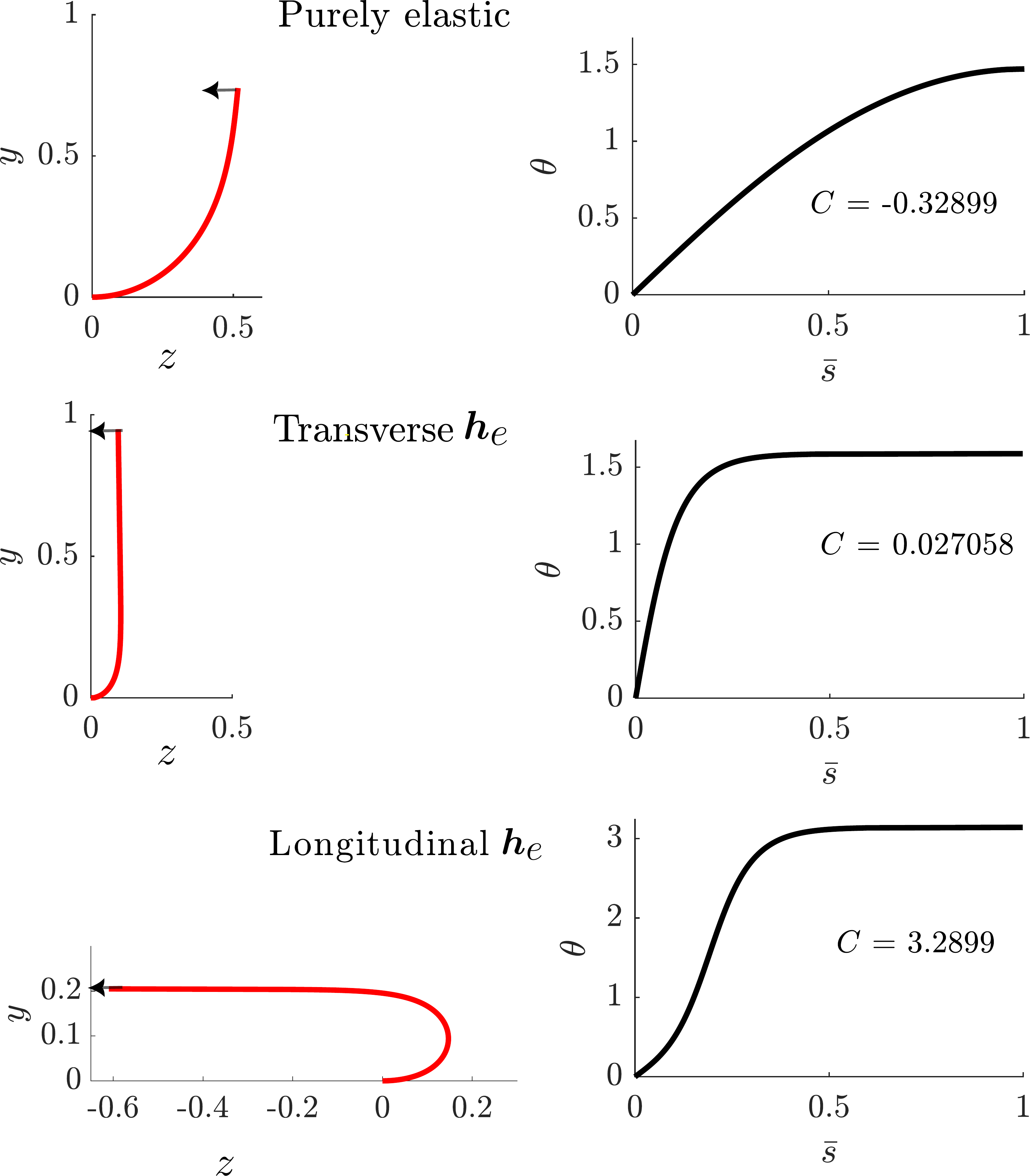}
	\caption{Deformed shapes of unit length Euler strut for purely elastic, transverse and longitudinal $\vb*{h}_e$ cases for $\bar{P} = \frac{\pi^2}{3}$ and $\bar{K}_d = 100$.}
	\label{fig:bvp-fixed-free-first-mode}
\end{figure}

We begin by obtaining the deformations corresponding to the first mode for the fixed-free case (Euler strut). The integration path starts from a point on the $\theta'$-axis ($\theta = 0$) for a trajectory $\mathcal{H}(\theta, \theta') = C$ close to the origin. We continue the integration along this trajectory until we intersect the $\theta$-axis ($\theta' = 0$), completing a quarter-revolution (see Fig. \ref{fig:schematic-trajectory}(a)). Next, we incrementally increase $C$ until the total arc length of the shape equals unity. The deformed shapes are illustrated in Fig. \ref{fig:bvp-fixed-free-first-mode}. The shapes corresponding to purely elastic and transverse $\vb*{h}_e$ cases for these canonical boundary conditions have been reported in \cite{AvatarDabade2024}.

To ensure a non-trivial buckled configuration of the Euler strut that satisfies the unit length criterion, we select $\bar{P} > \bar{P}_{\text{critical}} = \frac{\pi^2}{4}$. Note that if $\bar{P} < \bar{P}_{\text{critical}}$, we cannot find any shape for which arc length is equal to 1. For the purely elastic and transverse $\vb*{h}_e$ cases, we consider trajectories within the separatrix enclosing the origin (see Figs. \ref{fig:free-standing-elastica} and \ref{fig:free-standing-transverse-kdbar-100}). For the longitudinal $\vb*{h}e$ case, we examine trajectories that lie between the first and second separatrices (see Fig. \ref{fig:free-standing-longitudinal-kdbar-100}). The shapes corresponding to boundary value problems in this case lie between the separatrices. We observe significant differences in the deformation of ferromagnetic rods exposed to magnetic fields compared to purely elastic cases. For instance, the nature of loading in purely elastic case is compressive, while that for longitudinal $\vb*{h}_e$ case is tensile.

To obtain higher mode shapes for the fixed-free case, we integrate through an increasing number of odd quarter-revolutions, while ensuring that the specified $\bar{P}$ for each mode is greater than the corresponding $\bar{P}_{\text{critical}}$ for that mode.

Using a similar methodology that satisfies the boundary conditions, we determine the deformed shapes for pinned-pinned ($\theta'(\bar{s} = 0) = 0, \theta'(\bar{s} = 1) = 0$) and fixed-fixed ($\theta(\bar{s} = 0) = 0, \theta(\bar{s} = 1) = 0$) boundary conditions. All the pinned-pinned shapes are symmetric, resulting in $\bar{R} = 0$. In contrast, only the odd-numbered modes of the fixed-fixed configurations exhibit symmetry. Please refer to \cite{AvatarDabade2024} to see even numbered mode for $\bar{R}\neq 0$ for purely elastic and transverse $\vb*{h}_e$ cases. This symmetry ensures that the integral constraint (Eqn. \ref{eqn:constraints}) is automatically satisfied. Please refer to Eqns. (24,26,58) in \cite{Bigoni2015} for further justification.

Using our approach, we successfully obtain deformed configurations of the ferromagnetic rod with $\bar{R} = 0$. However, we are unable to capture the intriguing even-numbered mode shapes for the fixed-fixed scenario, which exhibit non-zero $\bar{R}$, as discussed in \cite{AvatarDabade2024,Antman1978}. This limitation underscores the complexity of these configurations and suggests potential areas for further investigation.

\begin{figure}[ht!]
	\centering
	\includegraphics[width=0.9\linewidth]{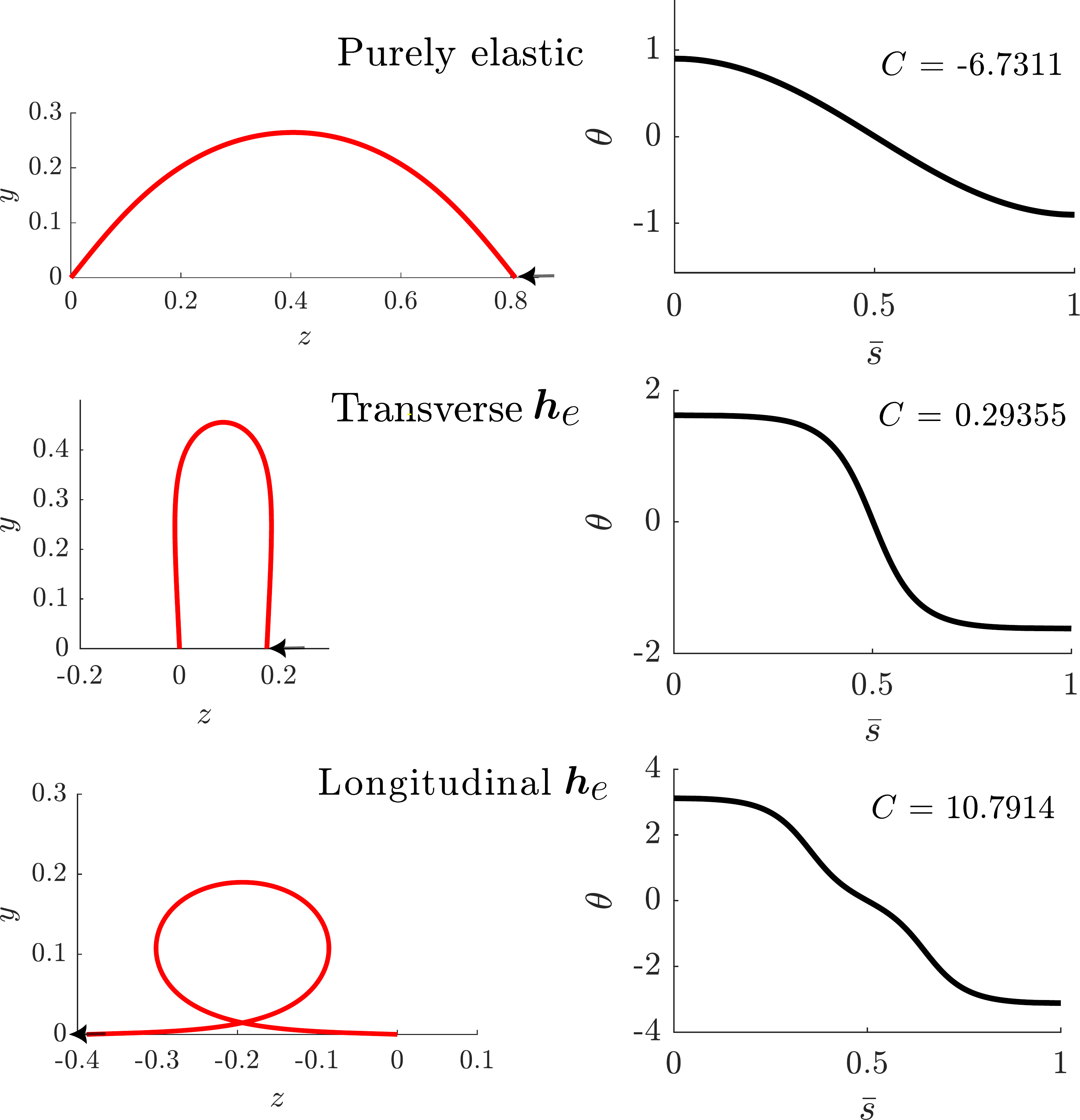}
	\caption{Deformed shapes of unit length pinned-pinned rod for purely elastic, transverse, and longitudinal $\vb*{h}_e$ cases at $\bar{P} = 1.1\pi^2$ and $\bar{K}_d = 100$.}
	\label{fig:bvp-pinned-pinned-first-mode}
\end{figure}

To obtain the first mode shape of a pinned-pinned configuration, we integrate along a trajectory starting from a point on the $\theta$-axis ($\theta' = 0$) until it intersects the $\theta$-axis again, thereby completing a half-revolution (see Fig. \ref{fig:schematic-trajectory}(b)). The deformed shapes are plotted in Fig. \ref{fig:bvp-pinned-pinned-first-mode} for $\bar{K}=100$ and $\bar{P} = 1.1\pi^2$, which exceeds the critical load for the purely elastic pinned-pinned rod, given by $\bar{P}_{\text{critical}} = \pi^2$. Note that, we observe a self-intersection in the rod when it is exposed to longitudinal $\vb*{h}_e$.

\begin{figure}[H]
	\centering
	\includegraphics[width=0.9\linewidth]{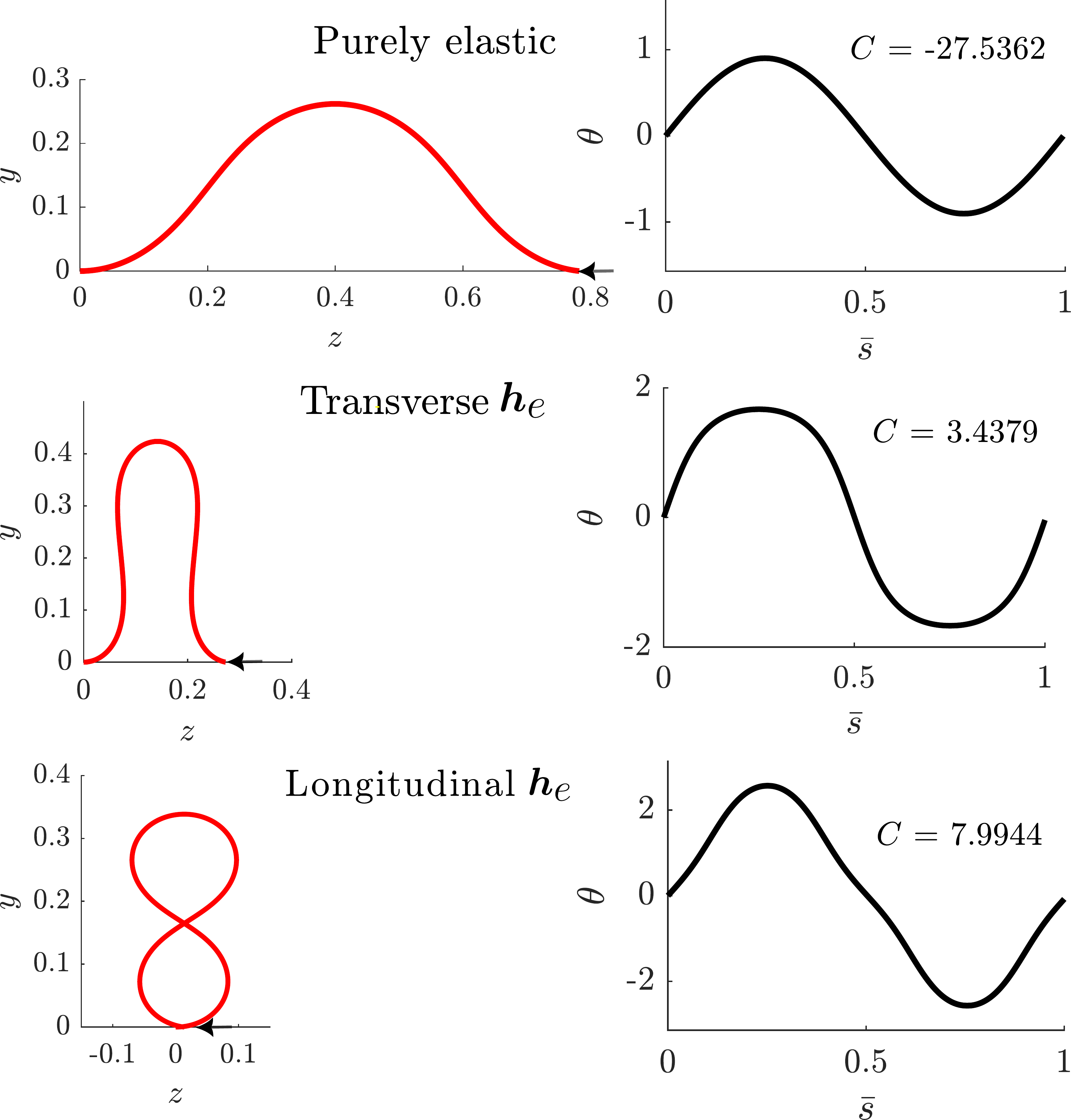}
	\caption{Deformed shapes of unit length fixed-fixed rod for purely elastic, transverse and longitudinal $\vb*{h}_e$ cases at $\bar{P} = 4.5\pi^2$ and $\bar{K}_d = 100$. }
	\label{fig:bvp-fixed-fixed-first-mode}
\end{figure}

To determine the first mode of the fixed-fixed rod, we start from a point on the vertical ($\theta'$)-axis and integrate along the trajectory until we return to the starting point, thus completing a full revolution (refer to Fig. \ref{fig:schematic-trajectory}(c)). We have chosen the value of $\bar{P}$ as $4.5\pi^2$, which is greater than the critical load for a purely elastic fixed-fixed rod ($\bar{P}_{\text{critical}} = 4\pi^2$). The resulting deformed shapes are shown in Fig. \ref{fig:bvp-fixed-fixed-first-mode}.

Determining the higher mode shapes of pinned-pinned configurations requires integrating over the corresponding number of half-revolutions for the appropriate value of $\bar{P}$. For the odd-numbered modes in the fixed-fixed case, we perform integration over the corresponding odd number of full revolutions. In all scenarios, the choice of the originating point ($\theta, \theta'$) must ensure that the total integration results in an arc length of unity.

\section{Conclusions}\label{sec:conclusion}
In this study, the equilibrium configurations of a ferromagnetic rod were analyzed under varying applied loads, while maintaining a fixed magnetization, utilizing the Kirchhoff kinetic analogy. Numerical methods were employed to compute the deformed configurations by leveraging the phase portrait trajectories of the corresponding Hamiltonian system. The numerical scheme was first validated against purely elastic rods, successfully reproducing the equilibrium configurations reported in \cite{Harmeet2018,Bigoni2015}, thereby establishing the reliability of the computational approach. Following this validation, we focused on a soft ferromagnetic rod under two distinct cases: one involving the application of a transverse external magnetic field and the other involving the application of a longitudinal external magnetic field. The external field was assumed to be sufficiently large to align the magnetization of the deformed ferromagnetic rod along the field direction. We showed that Kirchhoff's kinetic analogy can be extended to the ferromagnetic Kirchhoff rod. Using this analogy, we determined equilibrium shapes for certain free-standing ferromagnetic elastica as well as for specific boundary value problems.

A standard linear bifurcation analysis of the equilibrium equations for the elastica reveals that a rod undergoes a supercritical pitchfork bifurcation near the critical compressive load \cite{AvatarDabade2024}. Similarly, linear analysis of the equilibrium equations for a ferromagnetic ribbon shows that near the critical load, the ribbon experiences a supercritical pitchfork bifurcation under a transverse external magnetic field and a subcritical pitchfork bifurcation under a longitudinal external magnetic field. These equilibrium equations can be derived from the first variation of the total energy of the ferromagnetic ribbon, where the energy density is represented by the Lagrangian.

{\color{black}
The equilibrium configurations of elastic structures are governed by the Euler–Lagrange equations derived from the energy functional. Solving these equations yields the equilibrium shapes of the system. A linear bifurcation analysis of the equilibrium equations for the elastica (Euler strut with fixed-free boundary conditions) indicates that a rod undergoes a supercritical pitchfork bifurcation when subjected to a compressive load near a critical threshold \cite{AntmanBook}. A similar linear stability analysis of the equilibrium equations for a ferromagnetic ribbon reveals that near the critical load, the ribbon exhibits different bifurcation behaviors depending on the direction of the applied external magnetic field. Under a transverse external magnetic field, the ribbon undergoes a supercritical pitchfork bifurcation at a critical load $\bar{P}_1=-20.01\pi^2$. In contrast, when subjected to a longitudinal external magnetic field, the system experiences a subcritical pitchfork bifurcation at a critical load $\bar{P}_2=20.51\pi^2$ \cite{AvatarDabade2024}.

In our article, we have analyzed the phase portrait of the Hamiltonian system describing a ferromagnetic ribbon subjected to an external load. Our investigation reveals that, under a transverse external magnetic field, the phase portrait exhibits a subcritical pitchfork bifurcation at a load $\bar{P}_3=-20.26\pi^2$ as $\bar{P}$ is reduced. In contrast, when a longitudinal external magnetic field is applied, the phase portrait  of the Hamiltonian displays a supercritical pitchfork bifurcation at a load $\bar{P}_4=20.26\pi^2$. Notably, the bifurcation loads identified in the Hamiltonian phase portrait analysis differ from the critical loads obtained through linear bifurcation analysis of the equilibrium equations. This distinction emphasizes the difference between the critical buckling loads $(\bar{P}_1,\bar{P}_2)$ determined from the equilibrium equations and the bifurcation loads $(\bar{P}_3,\bar{P}_4)$ observed in the phase portrait of the Hamiltonian system.}

This work can be extended to analyze ferromagnetic elastic strips or ribbons by utilizing Kirchhoff's kinetic analogy. The Kirchhoff kinetic analogy can be suitably modified to account for the unique mechanics of elastic ribbons. Building on prior analyses, we aim to explore how the divergence of magnetization in ferromagnetic ribbons depends on their twist, \cite{AvatarDabade2024}. Future investigations will focus on studying the combined effects of twist and tension in ferromagnetic ribbons. Such an analysis is particularly compelling, as twist and tension in purely elastic rods are known to produce localized solutions, such as plectonemes \cite{Clauvelin2009}. Understanding how these phenomena manifest in ferromagnetic ribbons would offer valuable insights into the interplay between magnetism, elasticity, and geometric constraints in slender structures.

\section{Acknowledgments}\label{sec:acknowledgments}
This paper is dedicated to Prof. Richard James, whose scholarly work has inspired us, and future generations, to pursue research at the intersection of mechanics, materials science, and mathematics. We would like to acknowledge the Indian Institute of Science Startup Grant and the Prime Minister’s Research Fellowship (PMRF ID: 0201857) for providing financial support for this research.

\bibliographystyle{asmejour}

\bibliography{JAM_Avatar_Dabade}

\end{document}